\documentclass[a4paper,11pt]{article}
\pdfoutput=1 % if your are submitting a pdflatex (i.e. if you have
\usepackage{jheppub} % for details on the use of the package, please
\usepackage[T1]{fontenc} % if needed

\usepackage{xcolor} 

\usepackage{graphicx}
\usepackage{ntheorem}
\usepackage{mathrsfs}
\usepackage{subfigure}
\usepackage{dcolumn} 

\usepackage{hyperref}
\usepackage{comment}
\usepackage{bbm}

\usepackage{amsmath}
\usepackage{amssymb}
\usepackage{longtable}
\usepackage{boldline}
\usepackage{braket}

\usepackage{siunitx}

\usepackage{placeins}

\usepackage{array}

\title{\boldmath Logarithmic, Fractal and Volume-Law Entanglement in a Kitaev chain with long-range hopping and pairing}

\author[a,b,c,1]{Andrea Solfanelli,\note{Corresponding author.}}
\author[a,b,d]{Stefano Ruffo}
\author[c,e]{Sauro Succi}
\author[f]{Nicol\`o Defenu}

\affiliation[a]{SISSA, via Bonomea 265, I-34136 Trieste, Italy}
\affiliation[b]{INFN, Sezione di Trieste, Via Valerio 2, 34127 Trieste, Italy}
\affiliation[c]{\mbox{Center for Life Nano-Neuro Science @ La Sapienza, Italian Institute of Technology, 00161 Roma, Italy}}
\affiliation[d]{\mbox{Istituto dei Sistemi Complessi,
	Via Madonna del Piano 10, I-50019 Sesto Fiorentino, Italy}}
\affiliation[e]{\mbox{Physics Department, Harvard University, Oxford Street 17, Cambridge, USA}}

\affiliation[f]{Institut f\"ur Theoretische Physik, ETH Z\"urich, Wolfgang-Pauli-Str. 27 Z\"urich, Switzerland}

\emailAdd{asolfane@sissa.it}

\abstract{Thanks to their prominent collective character, long-range interactions promote information spreading and generate forms of entanglement scaling, which cannot be observed in traditional systems with local interactions. In this work, we study the asymptotic behavior of the entanglement entropy for Kitaev chains with long-range hopping and pairing couplings decaying with a power law of the distance. We provide a fully-fledged analytical and numerical characterization of the asymptotic growth of the ground state entanglement in the large subsystem size limit, finding that the truly non-local nature of the model leads to an extremely rich phenomenology. Most significantly, in the strong long-range regime, we discovered that the system ground state may have a logarithmic, fractal, or volume-law entanglement scaling, depending on the value of the chemical potential and on the strength of the power law decay.}

\begin{document} 
	\maketitle
	\flushbottom

	%\author{Andrea Solfanelli}
	%\email{asolfane@sissa.it}
	%\affiliation{SISSA, via Bonomea 265, I-34136 Trieste, Italy}
	%\affiliation{INFN, Sezione di Trieste, Via Valerio 2, 34127 Trieste, Italy}
	
	%\author{Stefano Ruffo}
	%\email{ruffo@sissa.it}
	%\affiliation{SISSA, via Bonomea 265, I-34136 Trieste, Italy}
	%\affiliation{INFN, Sezione di Trieste, Via Valerio 2, 34127 Trieste, Italy}
	%\affiliation{\mbox{
	%		Istituto dei Sistemi Complessi,
	%		Via Madonna del Piano 10, I-50019 Sesto Fiorentino, Italy
	%}}

    %\author{Sauro Succi}
    %\affiliation{\mbox{
    %		Center for Life Nano Science @ La Sapienza, Italian Institute of Technology, 00161 Roma, Italy}}
    %\affiliation{\mbox{Physics Department, Harvard University, Oxford Street 17, Cambridge, USA}}
    %

	%\author{Nicol\`o Defenu}
	%\email{ndefenu@phys.ethz.ch}
	%\affiliation{Institut f\"ur Theoretische Physik, ETH Z\"urich, Wolfgang-Pauli-Str. 27 Z\"urich, Switzerland}

%\date{\today} 
%
%
\section{Introduction}
In recent years, the quantum community's interest in long-range physics has steadily increased due to the emergence of promising platforms for quantum technological applications: long-range interacting quantum systems. These systems are characterized by coupling energies between pairs of microscopic constituents $V_{i,j}$ that decay as a power law of their distance $r=|i-j|$, with $\alpha>0$ \cite{Campa2014, DefenuArXiv2021}. This increased interest is largely due to the systems' stability against external perturbations, which allows for the mitigation of the detrimental effects of dynamically generated excitations \cite{DefenuArXiv2021, xuPhysics2022}. An example of the rigidity of long-range interacting platforms against external drivings and of its utility for quantum technological applications is the possibility for such systems to host clean discrete Floquet time crystal phases \cite{RussomannoPRB2017,SuracePRB2019,PizziNatComm2021,GiachettiArXiv2022}. Another example is the recently introduced advantage in the finite time performance of quantum heat-engines with a working substance hosting long-range couplings \cite{Solfanelli2022ArXiv}. 
Moreover, this technological and theoretical interest is also supported from the experimental side by the possibility to implement long-range interacting systems in typical quantum simulation platforms, such as atomic molecular and optical (AMO) systems \cite{BrittonNature2012,MonroeRevModPhys2021,MottlScience2012,MivehvarAdvancesinPhysics2021,ChomazArXiv2022}. Interestingly, trapped ions setups allow tuning the power law exponent $\alpha$, dictating the decay of the interaction energy with distance, from $\alpha \simeq 0$ to $\alpha \simeq 3$ \cite{BrittonNature2012}. 

The most important feature a system should have to be a good candidate for quantum technologies is the capability of hosting highly entangled states in its spectrum. Indeed, this crucial property is the essential ingredient to perform tasks that are classically impossible or very inefficient \cite{Bennett2000Nature}. More precisely, entanglement is the property that makes quantum computation overtake classical one providing the computational speed-up in quantum algorithms as compared to algorithms based on the processes of classical physics \cite{Nielsen2010}. Moreover, it is crucial for many quantum technological applications such as quantum teleportation \cite{Ma2012Nature}, quantum cryptography \cite{GisinRevModPhys2002} or quantum metrology \cite{GiovannettiNaturePhoton2011}.

A set of key quantities entering the characterization of entanglement is provided by the entanglement Rényi entropies. For their definition, one takes a partition of a given system into two subsystems $A$ and $B$ (the complement of $A$), determines the reduced density matrix of a subsystem (say, of $A$) $\rho_A$ by tracing out the degrees of freedom of $B$, and then computes its Rényi entropies: $S_{\nu} = \ln\mathrm{Tr}[\rho_A^\nu]/(1-\nu)$ \cite{Horodecki2009RMP}. One of the most fundamental properties of entanglement Rényi entropies is their behavior with the size of the subsystem considered. The celebrated area law \cite{Srednicki1993PRL,Eisert2010RevModPhys} refers to the fact that typically entanglement grows as the boundary of the subsystem considered, i.e., for a system in $d$ dimensions and a subsystem of size $L$ having volume $\sim L^d$ and area $\sim L^{d-1}$, then the entanglement entropy of the subsystem scales as $\sim L^{d-1}$. In particular, the area law has been proven to be satisfied in the ground state of one-dimensional systems with mass gap and short-range couplings when the size of the subsystem is much larger than the correlation length \cite{Hastings2007JStatMech}. At a quantum critical point, where the correlation length diverges,  the area law is known to be violated by a logarithmic term proportional to the central
charge of the conformal field theory (CFT) that describes the low-energy spectrum of the model \cite{Calabrese2004JStatMech, Calabrese2009JPA, Holzhey1994NPB, Vidal2003PRL, Latorre2004QuantumInfComp, Latorre2009JPA}. These facts motivated initially the study of this quantity due to its similarity to the black hole entropy \cite{Bombelli1986PRD,Srednicki1993PRL}, and have eventually revealed the important role that entanglement plays in high-energy physics \cite{Dong2016NatComm,Ryu2006PRL,VanRaamsdonk2010GenRelGrav,Maldacena2013FortschrPhys} as well as in the investigation of condensed matter systems \cite{Amico2008RevModPhys,Calabrese2009entanglement,Laflorencie2016PhysRep}.
 
The previous discussion changes and becomes more involved for systems with long-range couplings \cite{DefenuArXiv2021,Gong2017PRL,Kuwahara2020NatComm}. Indeed the prominent collective character of such non-local systems promotes entanglement spreading and leads to novel forms of equilibrium and dynamical scaling, which cannot be observed in traditional systems with local interactions \cite{Pappalardi2018PRB,Lerose2020PRR,Lerose2020PRA,GiachettiArXiv2021,Capizzi2022PRB,Pappalardi2019JStatMech}. In particular, the anomalous scaling of entanglement in the presence of long-range couplings has recently attracted great interest in the context of the so-called measurement-induced transitions \cite{Li2018PRB,Skinner2019PhysRevX,Chan2019PRB,Cao2019PRB,Gullans2020PRX,Turkeshi2021PRB,Alberton2021PRL}. In this case, the dynamical generation of entanglement is weakend by the presence of local measures applied randomly during the system evolution. More precisely, if the measurement rate is high enough, the steady state entanglement saturates to an area law value independent of the considered subsystem size, if only nearest neighbor interactions are present \cite{xuPhysics2022}. On the other hand, in the presence of long-range couplings, subvolume law scalings \cite{xuPhysics2022,Minato2022PRL,Block2022PRL,Muller2022PRL,Sharma2022SciPost}, also referred to as fractal entanglement phases \cite{Zhang2022Quantum,Ippoliti2022PRX}, appear. 

These interesting dynamical phenomena have no clear equilibrium counterpart showing that their origin is directly related to the presence of long-range interactions. The entanglement properties of the ground state of a fermionic chain with long-range pairing couplings and nearest neighbors hopping amplitudes were fully characterized in Refs. \cite{VodolaNJP2015,VodolaPRL2014,Ares2015PRA,Ares2018PRA,Ares2022JStatMech} which reported standard logarithmic violations of the area law in the weak long-range regime. Moreover, an anomalous logarithmic growth was found even if the mass gap is not zero, associated to the divergence of unnormalized couplings, in the strong long-range regime characterized by a power law decay exponent smaller than the system dimension. On the other hand, the authors of Refs. \cite{Gori2015PRB,Lepori2022PRR} considered a model of fermions with strong long-range hopping amplitudes and no pairing discovering a volume law entanglement scaling. Moreover, the entanglement properties of the Sachdev-Ye-Kitaev (SYK) model \cite{Sachdev1993PRL,Maldacena2016PRD}, i.e. a fully connected fermionic model with random interactions, have been extensively studied \cite{Zhang2022FrontPhys}. Also in this case, the eigenstates of the SYK Hamiltonian display a volume law entanglement scaling whose coefficient has been computed numerically using exact diagonalization techniques \cite{Liu2018PRB,Fu2016PRB} and analytically assuming the eigenstate thermalization hypothesis \cite{Huang2019PRD} or using a path-integral approach which becomes exact in the large-N limit \cite{Zhang2020SciPost,Haldar2020PRR}. Finally, also in long-range bosonic systems \cite{GhasemiNezhadhaghighi2012EPL,Sharma2022PRL} and in fully connected spin systems \cite{Latorre2005PRA,Vidal2007JStatMech,Orus2008PRL,Carrasco2016,Kumari2022Quantum} only logarithmic violations of the area law were reported. 

Despite the extensive amount of literature on the topic summarized above, none of the considered long-range models display a fractal entanglement scaling at equilibrium unless additional ingredients are added such as modifications of the couplings which violate time translational symmetry or the presence of a fractal Fermi surface\,\cite{Gori2015PRB}. Here, we are going to show that the subvolume law observed in measurement induced transitions\,\cite{xuPhysics2022,Minato2022PRL,Block2022PRL,Muller2022PRL,Sharma2022SciPost,Zhang2022Quantum,Ippoliti2022PRX} is directly caused by long-range interactions and also appears at equilibrium, provided certain conditions are met. 

To prove our claim, we study the ground state entanglement scaling in a prototypical model of fermions with power-law decaying hopping and pairing amplitudes, also known as the long-range Kitaev chain \cite{KitaevUFN2001,DefenuArXiv2021}. This model is sufficiently simple to allow us to perform analytic calculations but at the same time it turns out to host an extremely rich phenomenology. Using the well-known Fisher-Hartwig expansion \cite{fisher1969AdvChemPhys,Basor1994LinAlgApp}, we were able to analytically determine the leading order dependence of the ground state entanglement on the subsystem size $L$  in the scaling limit of an infinite chain of $N\to\infty$ sites and infinite subsystem $L\to\infty$ with fixed $l = L/N$, for different values of the available parameters. In particular, we can distinguish two main regimes: the weak long-range regime in which the coupling's power law decaying exponents are larger than the system dimension and the strong long-range regime in which they are smaller. In the former case, the system shows standard logarithmic deviations from the entanglement area law in correspondence with the quantum critical points, however, in the most interesting case of equal long-range hopping and pairing the coefficients in front of these logarithmic divergences show a nontrivial dependence on the power law decay exponent $\alpha$ which is not compatible with the standard scaling predicted by critical conformal field theory \cite{Calabrese2009JPA,Calabrese2004JStatMech}. On the other hand, in the strong long-range case, the system becomes genuinely non-additive, therefore showing a logarithmic deviation from the area law even away from criticality. Most significantly, when the system chemical potential is zero, no local terms are present in the Hamiltonian  (as we will see this simple fact strongly affects the nature of the ground state which becomes highly degenerate) thus resulting into a subvolume law entanglement scaling, $S\sim L^{1-2\alpha}$. 

Summarizing, our work correctly reproduces previously known results in different limits, thus bringing several disparate results present in the literature into a coherent picture. Moreover, we are able to detect a fractal entanglement scaling phase which is entirely due to the non-additive nature of the model and does not need the dynamical setting of measurement induced transitions to be observed.

The paper is organized as follows. In Section \ref{sec: Kitaev chain with long-range couplings} we introduce the long-range Kitaev model and we describe its phase diagram. In Section \ref{sec: Entanglement scaling} we briefly review the techniques which allow us to study the entanglement scaling of generic quadratic fermionic models (the expert reader may safely skip this part). Finally, Section \ref{sec: Weak long-range regime} and \ref{sec: Strong long-range regime} are devoted to the detailed characterization of the ground state entanglement scaling of the model in the weak and strong long-range regimes, respectively. 
\section{Kitaev chain with long-range couplings}\label{sec: Kitaev chain with long-range couplings}
We consider a generic model of spinless fermions hopping across the $N$ sites of a one-dimensional chain in the presence of pairing interactions, and with a chemical potential $h$. Assuming periodic boundary conditions, the system Hamiltonian reads
\begin{align}
	H = &-\sum_{j=1}^N\sum_{r=1}^{N/2-1}\left[t_r \hat{c}_{j+r}^\dagger \hat{c}_j+\Delta_r\hat{c}_{j+r}^\dagger \hat{c}_j^\dagger +h.c.\right]\notag\\
	&-h\sum_{j=1}^N\left[1-2\hat{c}_j^\dagger \hat{c}_j\right],\label{eq: long-range Kitaev chain H}
\end{align}
where $\hat{c}_j^\dagger$ and $\hat{c}_j$ are creation and annihilation operators for fermions at site $j$, while $t_r$ and $\Delta_r$ are the hopping and pairing amplitudes, respectively. We choose their dependence on the intersite distance $r$ according to the power laws
\begin{align}
	t_r = \frac{1}{N_{\alpha_1}}\frac{J}{r^{\alpha_1}},\quad \Delta_r = \frac{1}{N_{\alpha_2}}\frac{\Delta}{r^{\alpha_2}},
\end{align}
with the hopping exponent $\alpha_1>0$, the pairing exponent $\alpha_2>0$,  and $N_\alpha = \sum_{r=1}^{N/2}r^{-\alpha}$ the Kac scaling factor \cite{KacJMP1963}, which guarantees extensivity of the energy in the case $\alpha_i<1$, with $i = 1,2$. This model, often referred to as long-range Kitaev chain \cite{KitaevUFN2001}, is emerging as a minimal model for the study of the effects of long-range couplings on a quantum system \cite{DefenuArXiv2021}. Indeed, its integrable nature makes it amenable to both analytical and numerical treatment.
Moreover, as observed in Refs. \cite{jaschke2017critical,vanderstraeten2018quasiparticle,defenu2019dynamical}, when the pairing and hopping power law decay exponents are equal $\alpha_{1} = \alpha_{2} = \alpha$ the model can be related to the quantum Ising model. In particular, in the short-range case with $\alpha\to\infty$, the relation becomes exact through the Jordan-Wigner mapping\,\cite{Fradkin2013}. 

The quadratic nature of the Hamiltonian~\eqref{eq: long-range Kitaev chain H} allows its exact diagonalization in Fourier space via the Bogolyubov transformation

\begin{align}
	\hat{c}_k = \cos\frac{\theta_k}{2}\hat{\gamma}_k +\sin\frac{\theta_k}{2}\hat{\gamma}_{-k}^\dagger,
\end{align}
where we have introduced the momentum space fermionic operators
\begin{align}
	\hat{c}_k = \frac{e^{-i\frac{\pi}{4}}}{\sqrt{N}}\sum_{j=1}^N e^{ikj}\hat{c}_j,
\end{align}
where $k = 2\pi n/N$, and $n$ is an integer such that $\lfloor-N/2\rfloor+1\leq n\leq \lfloor N/2\rfloor$. While the Bogoliubov angles are defined by the conditions $\tan\theta_k = \tilde{\Delta}_k/(h-\tilde{t}_k)$, where Fourier transforms of the hopping and pairing amplitudes are defined as
\begin{align}
	\tilde{t}_k = \frac{J}{N_{\alpha_{1}}}\sum_{r=1}^{N/2-1}\frac{\cos(kr)}{r^{\alpha_{1}}},\quad
	\tilde{\Delta}_k = \frac{\Delta}{N_{\alpha_{2}}}\sum_{r=1}^{N/2-1}\frac{\cos(kr)}{r^{\alpha_{2}}}.\label{eq: hopping,pairing}
\end{align}
Hereafter, we set $J = \Delta = 1$ as the energy scale and work in units of $\hbar = 1$. In terms of the Bogoliubov fermions, the Hamiltonian then takes the diagonal form
\begin{align}
	H = \sum_k \omega_k(h)\left(\hat{\gamma}_k^\dagger\hat{\gamma}_k-1/2\right),
\end{align}
with the quasiparticle spectrum
\begin{align}
	\omega_k(h) = 2\sqrt{(h-\tilde{t}_k)^2+\tilde{\Delta}_k^2}.\label{eq: spectrum}
\end{align}
Since $\omega_k(h)\geq 0$, the ground state corresponds to the Fock space vacuum for the Bogoliubov modes, defined by the condition $\hat{\gamma}_k|\mathrm{gs}\rangle = 0, \forall k$.

\begin{figure*}
	\centering
	\includegraphics[width=\linewidth]{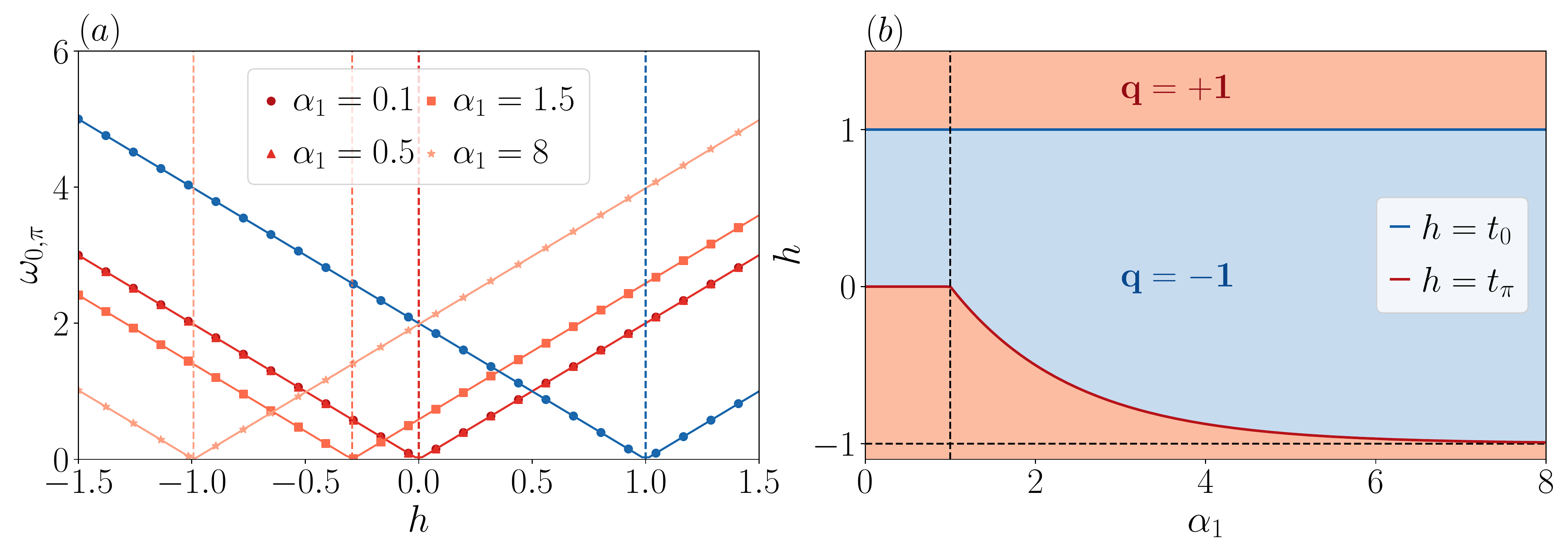}
	\caption{a) Critical modes $k = 0,\pi$ of the quasiparticle spectrum as a function of the chemical potential $h$ for different values of $\alpha_{1} = \alpha_{2}$, two critical points emerge at $h = \tilde{t}_0,\tilde{t}_\pi$, where in the thermodynamic limit $\tilde{t}_0 = 1$ and $\tilde{t}_\pi = 1$ if $\alpha_{1}>1$, $\tilde{t}_\pi = -1+2^{1-\alpha_{1}}$ if $1<\alpha_{1}<2$, and $\tilde{t}_\pi = 0$ if $0<\alpha_{1}<1$. b) Phase diagram of the long-range Kitaev chain in the plane $(\alpha_1,h)$, for the pairing decay exponent $\alpha_2= \alpha_1$, $\alpha_1$ is the hopping decay exponent and $h$ is the chemical potential. The topological order parameter is $q=-1$ in the topological phase (blue shaded region) and $q=+1$ in the trivial phase (red shaded region). The phase space boundaries correspond to the solid lines $h = \tilde{t}_0$ and $h = \tilde{t}_\pi$.}
		%Topological order parameter $q = -1$ in the topological phase (blue shaded region) and $q = +1$ in the trivial phase (red shaded region)plotted as a function of the hooping decay exponent $\alpha_{1}$ and of the chemical potential $h$. The phase space boundaries correspond to the solid lines $h = \tilde{t}_0$ and $h = \tilde{t}_\pi$.}
	\label{fig: kc_hc}
\end{figure*}

When studying the critical properties associated with the spectrum~\eqref{eq: spectrum}, we must distinguish two main regimes: the weak long-range regime when $\alpha_1,\alpha_2>1$, i.e., the power law decay exponents are larger than the system dimensionality, and the strong long-range regime when $\alpha_1,\alpha_2< 1$. In the weak long-range case, the Kac scaling is a constant in the thermodynamic limit: $N_{\alpha>1}\to \zeta(\alpha)$, where $\zeta(\alpha)$ is the Riemann zeta function. Moreover, when the system size goes to infinity, we can safely perform a continuum limit in the $k$ variable. In particular, Eq.~\eqref{eq: hopping,pairing} may be written as
\begin{align}
	\tilde{t}_k = \mathrm{Re}\left[\mathrm{Li}_{\alpha_1}(e^{ik})\right]/\zeta(\alpha_1),\quad
	\tilde{\Delta}_k = \mathrm{Im}\left[\mathrm{Li}_{\alpha_2}(e^{ik})\right]/\zeta(\alpha_2),
\end{align}
where $\mathrm{Li}_{\alpha}(z)$ denotes the polylogarithm function. This leads to a continuum spectrum $\omega_k$ characterized, at the critical points, by a dispersion relation that depends on $\alpha_1$ and $\alpha_{2}$. In particular, for $\alpha_1,\alpha_2>1$, the system possesses two different phases separated by two quantum critical points $h_{c} =1,-1+2^{1-\alpha_1}$, in correspondence of which the dispersion relation becomes gapless near to the critical mode $k_c = 0,\pi$, respectively \cite{UhrichPRB2020,DefenuArXiv2021}. The critical modes of the spectrum are shown in Fig. \ref{fig: kc_hc}a where $\omega_{0(\pi)}$(blue(red) lines in the plot) is plotted as a function of $h$ for different values of $\alpha_1 = \alpha_{2}$. The nature of the transition is topological and the two topological phases can be distinguished by the value of the bulk topological invariant \cite{Alecce2017PRB}
\begin{align}
	w = \oint\frac{d\theta_k}{2\pi} = \begin{cases}
		1 &\mathrm{if}\quad h\in [-1+2^{1-\alpha_{1}},1]\\
		0 &\mathrm{otherwise}
	\end{cases},\label{eq: winding number}
\end{align}
where the Bogoliubov angles are defined as $\theta_k = \arctan(\tilde{\Delta}_k/(h-\tilde{t}_k))$. Moreover, in the nontrivial phase with $w = 1$, the ground state is
doubly degenerate, and can support Majorana edge modes \cite{Jager2020PRB}. 

In the strong long-range regime $0<\alpha_1,\alpha_2<1$ the scenario is more complicated. Indeed, in this case, the Kac normalization factor $N_\alpha$ diverges at large $N$ as $N_\alpha\approx N^{1-\alpha}$, and the thermodynamic limit of Eq.~\eqref{eq: hopping,pairing} has to be carefully considered. In particular, as pointed out in Ref. \cite{DefenuPNAS2021}, while the Fourier modes variable $k = 2\pi n/N$ becomes continuous as $N\to\infty$, the hopping and pairing amplitudes $\tilde{t}_k$, $\tilde{\Delta}_k$, remain discrete and labeled by the integer $n$, reading
\begin{align}
	\lim_{N\to\infty} \tilde{t}_k &= c_{\alpha_{1}}\int_0^{1/2}ds\frac{\cos(2\pi n s)}{s^{\alpha_{1}}}= \tilde{t}_n,\\
	\lim_{N\to\infty} \tilde{\Delta}_k  &= c_{\alpha_{2}}\int_0^{1/2}ds\frac{\sin(2\pi n s)}{s^{\alpha_{2}}}= \tilde{\Delta}_n,
\end{align}
with $c_\alpha = (1-\alpha)2^{1-\alpha}$. Therefore, the presence of long-range couplings leads to a discrete spectrum $\omega_k\to\omega_n = 2\sqrt{(h-\tilde{t}_n)^2+\tilde{\Delta}_n^2}$ also at $N\to\infty$. The persistence of the discrete spectrum in the thermodynamic limit does not allow us to define a continuous theory and hinders the conventional definition of quantum critical points in the Kitaev chain. In particular, the winding number in Eq.~\eqref{eq: winding number} is ill-defined as a consequence of the discontinuity in the Bogolyubov angle distribution \cite{Alecce2017PRB}. Yet, the transition can still be characterized by the quantity
\begin{align}
	q = \mathrm{sign}[(h-\tilde{t}_0)(h-\tilde{t}_\pi)] = \begin{cases}
		1 &\mathrm{if}\quad h\in [\tilde{t}_\pi,\tilde{t}_0]\\
		-1 &\mathrm{otherwise}
	\end{cases}.
\end{align}
This quantity has proven to be a good topological invariant in cases in which the winding number turns out to be ill-defined \cite{Alecce2017PRB,DeGottardi2013PRB}. Then, also in the strong long-range regime, the behavior of the order parameter $q$ is still consistent with a change of phase at the critical points $h = \tilde{t}_0,\tilde{t}_\pi$ \cite{DefenuPNAS2021}. %, which corresponds to the points at which the $k = 0,\pi$ modes of the bulk spectrum have zero energy \cite{DefenuPNAS2021}.
However, as shown in \cite{lepori2017NJP}, the bulk boundary correspondence turns out to be weakened by the presence of strong long-range couplings. Consequently, the change of $q$ at the critical points is not guaranteed to be in one-to-one correspondence with the appearance of boundary topological edge states. Nevertheless, we expect bulk properties to remain consistent with a change of phase. Figure \ref{fig: kc_hc}b shows the model phase diagram as characterized by the value of $q = \pm 1$ as a function of the chemical potential $h$ and of the hopping power law decay exponent $\alpha_{1}$. Two quantum critical lines appear when varying the $\alpha_1$ parameter. In particular, we notice that the location of the critical point corresponding to $\omega_0 = 0$ is fixed to $h = \tilde{t}_0 = 1$ for any value of $\alpha_{1}$ (blue bold line in Fig. \ref{fig: kc_hc}b). On the contrary, the critical point corresponding to $\omega_\pi = 0$ (red bold line in Fig. \ref{fig: kc_hc}b) is $\alpha_{1}$ dependent with two different behaviors in the weak and strong long-range regimes, in particular in the thermodynamic limit we find 
\begin{align}
	\lim_{N\to\infty} \tilde{t}_{\pi} = \begin{cases}
		-1&\mathrm{if}\quad \alpha_1>2\\
		-1+2^{1-\alpha_{1}}  &\mathrm{if}\quad 1<\alpha_1<2\\
		0 &\mathrm{if}\quad 0<\alpha_1<1
	\end{cases}.
\end{align}

Finally, the completely mean-field case with $\alpha_1 = \alpha_{2} = 0$ needs to be treated separately. Indeed, in this case, the spectrum becomes strongly degenerate and this may alter the nature of the ground state. In particular, for completely flat couplings the sums in Eq.~\eqref{eq: hopping,pairing} can  be exactly computed and, in the thermodynamic, they read 
\begin{align}
	\tilde{t}_n(\alpha_{1} = 0) =\delta_{n,0}, \quad
	\tilde{\Delta}_n(\alpha_{2} = 0) = \frac{1+(-1)^{n+1}}{\pi n}.
\end{align}
Accordingly, the single-particle spectrum becomes
\begin{align}
	\omega_{n}^0 = \begin{cases}
		2|h|                                 &\mathrm{if}\,|n|\,\mathrm{even}
		\\
		2\sqrt{h^2+4/(\pi n)^2} &\mathrm{if}\,|n|\,\mathrm{odd}
		\\
		2|h-1|                               &\mathrm{if}\,\,n=0
	\end{cases},
\end{align}
where we have introduced the shortcut notation $\omega_{n}^0 = \omega_{n}(\alpha_{1} = 0,\alpha_{2} = 0)$. It follows that an extensive number of single-particle energy levels corresponding to all the even modes become degenerate. In particular, when the chemical potential is zero $h = 0$ all the even modes become zero modes since at this point we have $\omega_{2n}^0(h = 0) = 0$, $\omega_{2n+1}^0(h = 0) = 2/|\pi n|$ and $\omega_0^0(h = 0) = 1$. This fact deeply affects the nature of the many-body ground state which is no more given by the Bogoliubov vacuum, on the contrary, it allows for a finite population of Bogoliubov fermions in an extensive number of zero modes. More precisely, the ground state for $\alpha_{1,2} = 0$ and $h = 0$ is given by a generic superposition of the form
\begin{align}
	|\mathrm{gs}\rangle_{\alpha = 0,h=0} = \sum_{n_0=0}^{N_0}C_{n_0}|n_0\rangle,
\end{align}
where $n_0$ is the number of fermions occupying the $N_0$ available zero modes. This ground state is highly degenerate indeed each $|n_0\rangle$ state can be realized in $\binom{N_0}{n_0}$ ways, leading to the exponential degeneracy 
\begin{align}
	\mathrm{Deg}[|\mathrm{gs}\rangle_{\alpha = 0,h=0}] = \sum_{n_0 = 0}^{N_0}\binom{N_0}{n_0} = 2^{N_0}.	
\end{align}

As a concluding remark for this section, we stress the importance of the Kac scaling in the stabilization of the topological order in the strong long-range regime. Indeed, had we considered not properly rescaled couplings, the presence of long-range hopping $\alpha_1<1$ would have moved the critical point to $h_c = \mathcal{O}(N^{1-\alpha_1})\to\infty$, thus destroying the transition.

\section{Entanglement scaling in free fermionic systems}\label{sec: Entanglement scaling}
\begin{figure*}
	\centering
	\includegraphics[width=0.45\linewidth]{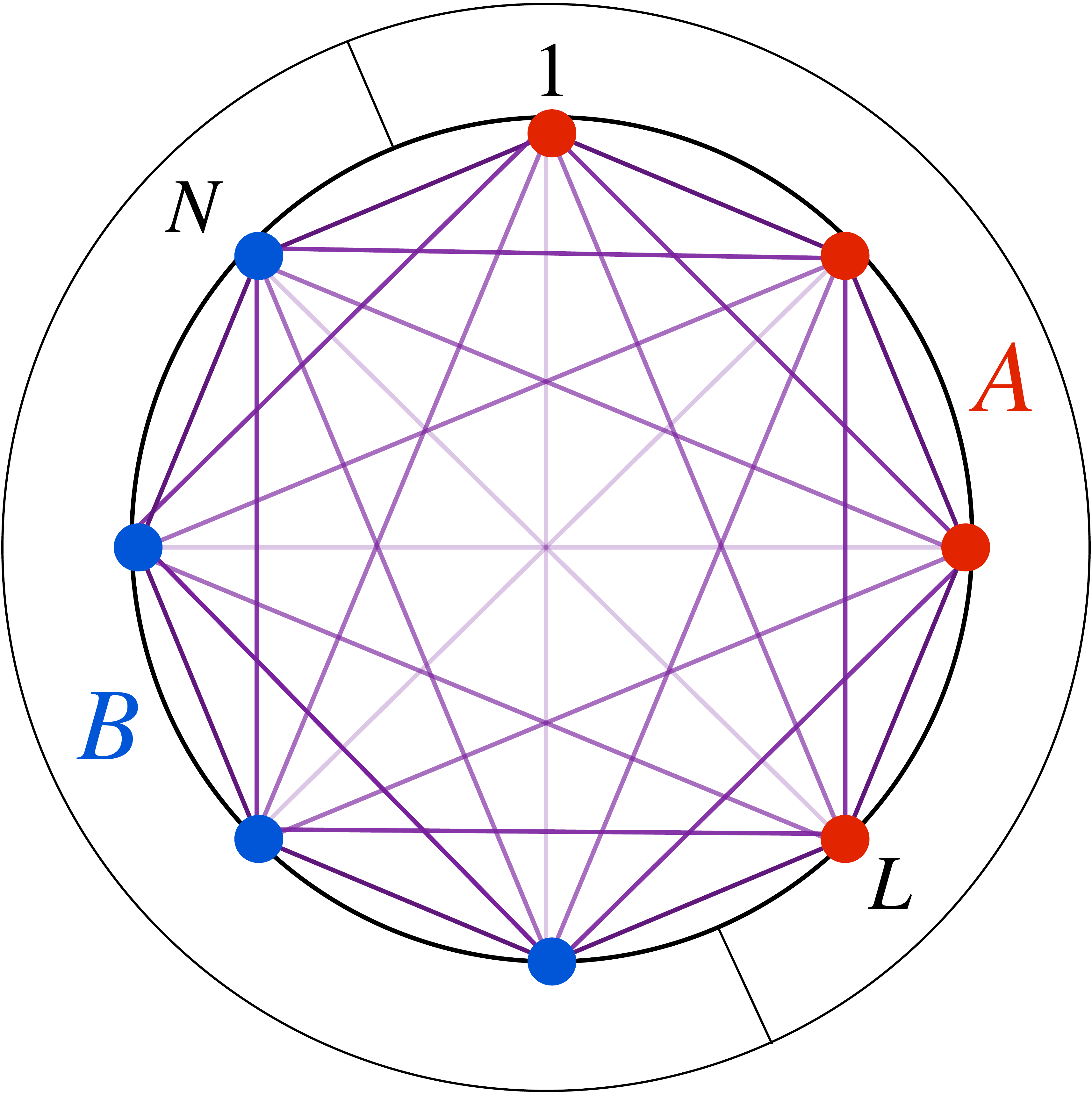}
	\caption{Schematic representation of a bipartition of a long-range Kitaev chain with periodic boundary conditions in two subsystems $A$ and $B$ of length $L$ and $N-L$ respectively.}
	\label{fig: Bipartition}
\end{figure*}
We consider a bipartition of the fermionic chain described by the Hamiltonian in Eq.~\eqref{eq: long-range Kitaev chain H}, in two subsystems $A$ and $B$, where $A$ is a continuous interval of chain sites of length $L$ and $B$ is its complementary set, see Fig. \ref{fig: Bipartition}. Given the Hilbert spaces $\mathcal{H}_A$ and $\mathcal{H}_B$ associated to $A$ and $B$, respectively, then the total Hilbert space of the system can be written as the tensor product $\mathcal{H} = \mathcal{H}_A\otimes\mathcal{H}_B$. If the total system is in a pure state $|\psi\rangle$, then the reduced density matrix, describing the state of subsystem $A(B)$ is obtained by taking the partial trace with respect to $\mathcal{H}_{A(B)}$: $\rho_{A(B)} = \mathrm{Tr}_{A(B)}|\psi\rangle\langle\psi|$. The amount of entanglement between the two subsystems can be characterized by the so-called Rényi entropies of $A$, defined as 
\begin{align}
	S_{\nu,L}(A) = \frac{1}{1-\nu}\ln\mathrm{Tr}[\rho_A^\nu],
\end{align}
where $\nu\geq 1$. These are known to provide an accurate measure for the entanglement of a bipartite system in a pure state \cite{Horodecki2009RMP}. In particular, the limit $\nu\to 1$ of the above expression corresponds to the celebrated Von Neumann or entanglement entropy 
\begin{align}
	S_L(A) = S_{1,L}(A) = -\mathrm{Tr}[\rho_A\ln\rho_A].
\end{align}
The main goal of this paper is to study the Rényi entanglement entropy for the ground state of a Hamiltonian of the kind analyzed in the previous Section. In particular, we are interested in determining the dependence of $S_{\nu,L}(A)$ on the subsystem size $L$ in the scaling limit $N\to\infty, L\to\infty$ with fixed $l = L/N$ and how this is affected by the presence of long-range hopping and pairing couplings in the Hamiltonian. This task may be achieved by taking advantage of the fact, that since the Hamiltonian in Eq.~\eqref{eq: long-range Kitaev chain H} is quadratic, then all its eigenstates satisfy the Wick decomposition theorem \cite{Vidal2003PRL,Peschel2003JPA}. Accordingly, the reduced density matrix can be obtained from the two-point correlation functions. To achieve this, we introduce the $2N\times 2N$ correlation matrix $\mathbb{V}$, which is a block matrix with each $2\times 2$ block defined as follows:
\begin{align}
	\mathbb{V}_{ij} = \begin{pmatrix}
		\delta_{ij}-2\langle c_j^\dagger c_i\rangle &2\langle c_i c_j\rangle
		\\
		2\langle c_i^\dagger c_j^\dagger\rangle &2\langle c_i^\dagger c_j\rangle-\delta_{ij}\
	\end{pmatrix},\label{eq: correlation matrix}
\end{align}
where $i$ and $j$ range from $1$ to $N$. Then, it can be shown \cite{Vidal2003PRL,Peschel2003JPA} that this is related to the Rényi entropies through the formula
\begin{align}
	S_{\nu,L}(A) = \frac{1}{2(\nu-1)}\mathrm{Tr}\ln\left[\left(\frac{\mathbb{I}+\mathbb{V}}{2}\right)^\nu+\left(\frac{\mathbb{I}-\mathbb{V}}{2}\right)^\nu\right].\label{eq: Renyi entropy formula}
\end{align}
It is important to notice that, from the computational point of view, this formula constitutes a dramatic simplification since the problem complexity is reduced from the diagonalization of a reduced density matrix of size $2^L\times 2^L$ to the diagonalization of the correlation matrix~\eqref{eq: correlation matrix} of size $2L\times 2L$, thus allowing to reach larger sizes $L$. From the analytic side, it is useful to write Eq.~\eqref{eq: Renyi entropy formula} as an integral on the complex plane along a contour $\mathcal{C}$ surrounding the
eigenvalues $v_j\in [-1,1]$ of $\mathbb{V}$. Using Cauchy's residue theorem in order to perform the integral, one gets \cite{jin2004JSP,its2008CommMathPhys}
\begin{align}
	S_{\nu,L}(A) = \lim_{\epsilon\to 0^+}\oint_{\mathcal{C}} s_\nu(1+\epsilon,z)\frac{d\ln D_L(z)}{dz}dz,\label{eq: Rényi integral}
\end{align}
where we have introduced the function
\begin{align}
	s_\nu(x,y)= \frac{1}{1-\nu}\ln\left[\left(\frac{x+y}{2}\right)^\nu+\left(\frac{x-y}{2}\right)^\nu\right],\label{eq: s_nu}
\end{align}
and the determinant 
\begin{align}
	D_L(z) = \det(z\mathbb{I}-\mathbb{V}).
\end{align}
Due to the translational invariance of the Hamiltonian~\eqref{eq: long-range Kitaev chain H} and given the choice of subsystem A, which is composed of contiguous sites, we can write the Fourier trasform of the correlation matrix $\mathbb{V}_{lj}$ as
\begin{align}
	\mathbb{V}_{lj} = \frac{1}{N}\sum_k G_k e^{ik(l-j)},
\end{align}
where we have introduced the two dimensional symbol $G_k$ which, as detailed in Appendix \ref{app: Derivation of the matrix symbol}, can be written as
%for the
%ground state of the Hamiltonian~\eqref{eq: long-range Kitaev chain H}, has the form
%
%\begin{align}
%	G_k = \frac{1}{\omega_k}\begin{pmatrix}
%		h-\tilde{t}_k &i\tilde{\Delta}_k\\
%		-i\tilde{\Delta}_k &\tilde{t}_k-h
%	\end{pmatrix}.
%\end{align}
\begin{align}
	G_k = (1-(f_k+f_{-k}))\left[\frac{2(h-\tilde{t}_k)}{\omega_k}\sigma_z-\frac{2\tilde{\Delta}_k}{\omega_k}\sigma_y\right]-(f_k-f_{-k})I,\label{eq: matrix symbol}
\end{align}
where $\sigma_a$, with $a = x,y,z$, are the Pauli sigma matrices, $I$ is the $2\times 2$ identity, and $f_k = \langle\hat{\gamma}_k^\dagger\hat{\gamma}_k\rangle$ are the occupation numbers of the Bogoliubov fermionic modes, which for a generic state satisfy the condition $0\leq f_k\leq 1$. 

Using the techniques introduced in Refs. \cite{Vidal2003PRL,Peschel2003JPA} the asymptotic behavior for $L\to\infty$ of the Toeplitz determinant $D_L(z)$, entering the expression for the Rényi entropies~\eqref{eq: Renyi entropy formula}, can be determined applying the Szegő-Widom theorem \cite{widom1974AdvMAth,widom1976AdvMAth} and an extension of the Fisher-Hartwig conjecture \cite{fisher1969AdvChemPhys,Basor1994LinAlgApp} to non-scalar symbols \cite{Ares2015PRA,Ares2018PRA}. The leading order contributions to the logarithm of $D_L(z)$ in the $L\to\infty$ limit then read
\begin{align}
	\ln D_L(z) &= \frac{L}{2\pi}\int_{-\pi}^\pi dk\ln\det(z\mathbb{I}-G_k)\notag
	\\
	&+\ln L\sum_{p}b_p(z)+\mathcal{O}(1),\label{eq: det expansion}
\end{align}
where the coefficients $b_p(z)$ of the logarithmic contribution are associated to the discontinuities of $G_k$. More precisely, if there is a discontinuity at some $k = p$, this means that 
\begin{align}
	G_p^+ =\lim_{k\to p^+}G_k\neq \lim_{k\to p^-}G_k = G_p^-,
\end{align}
then the coefficient corresponding to such discontinuity can be computed as \cite{Ares2018PRA} 
\begin{align}
	b_p(z) = \frac{1}{4\pi^2}\mathrm{Tr}[\ln (z\mathbb{I}-G_p^-)(z\mathbb{I}-G_p^+)^{-1}]^2.\label{eq: b coefficients}
\end{align}
Inserting Eq.~\eqref{eq: det expansion} into the integral for the Rényi entropy~\eqref{eq: Rényi integral} one obtains
\begin{align}
	S_{\nu,L} = \frac{1}{1-\nu}\sum_k\ln\left[(1-f_k)^\nu+f_k^\nu\right] +B_{\nu}\ln L +\mathcal{O}(1),\label{eq: Fisher-Hartwig expansion}
\end{align}
where the coefficient of the logarithmic contribution can be computed as
\begin{align}
	B_{\nu} = \sum_p\lim_{\epsilon\to 0^+}\oint_{\mathcal{C}} s_\nu(1+\epsilon,z)\frac{d b_p(z)}{dz}dz.
\end{align}
As shown in Section \ref{sec: Kitaev chain with long-range couplings}, whenever $\alpha_{1,2}>0$ or $\alpha_{1}=\alpha_{2} = 0$ and $h\neq 0$, the many-body ground state of the system is the Bogoliubov vacuum with $f_k = 0$ $\forall k$, therefore we are left with a leading order contribution given by a constant term $\mathcal{O}(1)$ corresponding to the standard area law in the one-dimensional case, or a logarithmic contribution which is associated to the discontinuity of the correlation matrix symbol $G_k$. On the other hand in the specific case $\alpha_{1}=\alpha_{2}=0$ and $h=0$ the many-body ground state becomes highly degenerate allowing for a finite fermionic population $f_k\neq 0$ for an extensive number of Bogoliubov modes, i.e., all the even modes. As a consequence, the first term in Eq.~\eqref{eq: Fisher-Hartwig expansion} becomes the leading contribution to the large $L$ entanglement scaling corresponding to a volume law behavior $S_{\nu,L}(\alpha_{1,2} = 0, h=0)\approx L$.

Summarizing, the machinery introduced in this section allows us to compute the leading order contribution to the scaling of Rényi entropies with the subsystem size by simply analyzing the symbol continuity properties in the different regimes. 
\section{Weak long-range regime}\label{sec: Weak long-range regime}
Let us start with the weak long-range regime corresponding to $1<\alpha_1,\alpha_2<2$. In this case, as we have seen in Section \ref{sec: Kitaev chain with long-range couplings}, the quasiparticle spectrum is continuous in the thermodynamic limit, and the ground state is always given by the Bogoliubov vacuum with zero fermionic populations $f_k = 0$, $\forall k$. Accordingly, the first term of the Fisher-Hartwig expansion~\eqref{eq: Fisher-Hartwig expansion} vanishes and then the leading order contribution to the entanglement scaling comes from the logarithmic term associated with the matrix symbol discontinuity. 
  
Within the weak long-range regime, we can distinguish three different cases: $\alpha_1>\alpha_2$, $\alpha_1<\alpha_2$ and $\alpha_1 = \alpha_2 = \alpha$. Therefore, in order to proceed we must identify the location of the jumps of $G_k$ and compute the lateral limits in these three different situations. Possible sources of discontinuities for $G_k$ are the discontinuities or the zeros of the spectrum $\omega_k(h)$, which appear at the two quantum critical points $h = 1, -1+2^{1-\alpha_{1}}$, where the spectrum becomes gapless at the soft modes $k = 0,\pi$, respectively. More precisely, $G_k$ has no discontinuities when $h\neq 1,-1+2^{1-\alpha_{1}}$, since in this case the lateral limits at the critical modes read
\begin{align}
	G_{0}^{\pm} &= \lim_{k\to 0^\pm} G_k = \mathrm{sgn}(h-1)\sigma_z,\\
	G_{\pi}^{\pm} &= \lim_{k\to \pi^\pm} G_k = \mathrm{sgn}(h+1-2^{1-\alpha_1})\sigma_z.
\end{align}
This leads to a constant scaling of the entanglement entropy $S_{\nu,L} = \mathcal{O}(1)$ with the subsystem size when the system is not at quantum criticality and therefore the spectrum is gapped. This is nothing but a manifestation of the standard area law for one-dimensional gapped systems \cite{Srednicki1993PRL,Eisert2010RevModPhys}. On the other hand, quantum criticality leads to logarithmic deviations from the area law. Let us start from the homogeneous critical point ($h = 1$), when the spectrum has an $\alpha_{1,2}$ dependent dispersion relation (see Appendix \ref{app: Dispersion relation around the critical modes}), which leads to the different lateral limits
\begin{align}
	G_{0}^{\pm} = \begin{cases}
		\sigma_z &\mathrm{if}\quad \alpha_1<\alpha_2
		\\
		-\sin(\alpha\pi/2)\sigma_z\pm\cos(\alpha\pi/2)\sigma_y &\mathrm{if}\quad \alpha_1=\alpha_2
		\\
		\pm\sigma_y &\mathrm{if}\quad \alpha_1>\alpha_2
	\end{cases},\label{eq: lateral limits h1wlr}
\end{align}
Accordingly, no discontinuity is present when the power law decay of the hopping amplitude is slower than that of the pairing, leading again to a constant entanglement entropy. In the $\alpha_1>\alpha_2$ case instead, we have a discontinuity in the symbol, with commuting lateral limits. Inserting the expression for $G_0^\pm$ in Eq.~\eqref{eq: b coefficients} we obtain
\begin{align}
	b_0(z) = \frac{1}{2\pi^2}\left(\ln\left(\frac{z+1}{z-1}\right)\right)^2.
\end{align}
Then, inserting this result into the expression for the entanglement entropy, and performing the integration in Eq.~\eqref{eq: Rényi integral} we obtain the logarithmic scaling
\begin{align}
	S_{\nu,L} = \frac{\nu+1}{12\nu}\ln L +\mathcal{O}(1).\label{eq: entanglement scaling c= 1/2}
\end{align}
\begin{figure*}
	\centering
	\includegraphics[width=\linewidth]{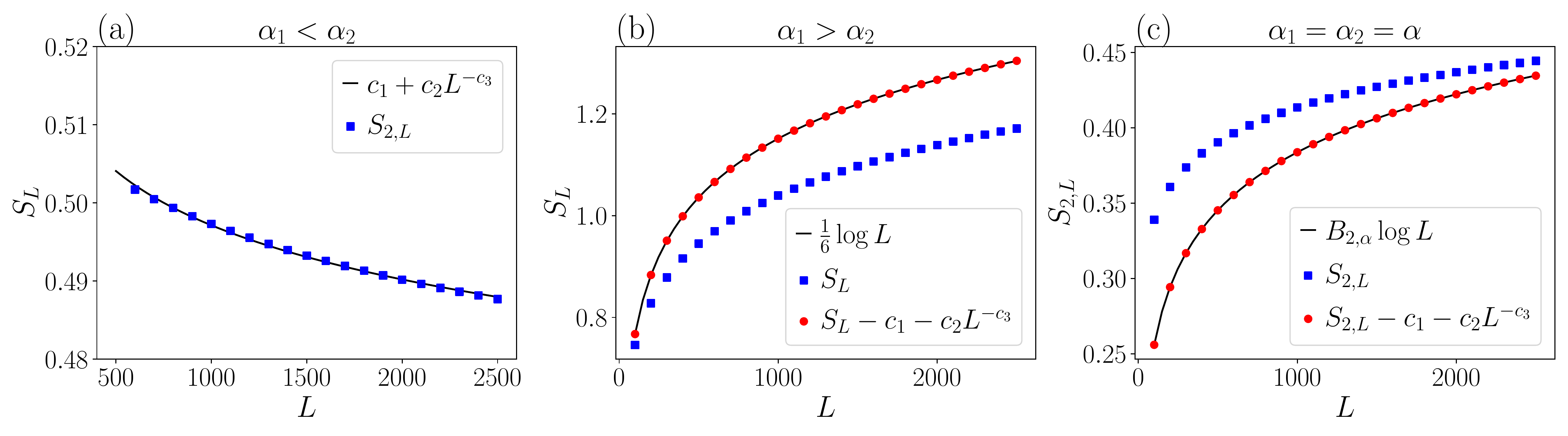}
	\caption{Numerical check of the entanglement scaling as a function of the subsystem size $L$ at the quantum critical point with chemical potential $h = 1$ for different values of couplings power law decay exponents $1<\alpha_1, \alpha_{2}$ and total system size $N = 2L$. a) Entanglement entropy ($\nu = 1$), with $\alpha_1 = 1.5$ and $\alpha_{2} = 1.8$, blue squares represent the numerical data while the black solid line is a fit of a constant and a subleading contribution $c_1+c_2L^{-c_3}$. b) Entanglement entropy ($\nu = 1$), with $\alpha_1 = 1.8$ and $\alpha_{2} = 1.5$, blue squares represents the numerical data, the black solid line correspond to the curve $(1/6))\ln L$, red dots have been obtained from the numerics by subtracting the fit of the subleading corrections of the form $c_1+c_2L^{-c_3}$. c) Rényi-2 entropy ($\nu = 2$) with $\alpha_1 = \alpha_{2} = 1.5$, blues squares represents the numerics, the black solid line represents the curve $B_{2,\alpha}\ln L$, red dots are obtained subtracting the subleading corrections to the numerical data as in panel b).}
	\label{fig: EntanglementScaling_hf}
\end{figure*}
\begin{figure*}
	\centering
	\includegraphics[width=\linewidth]{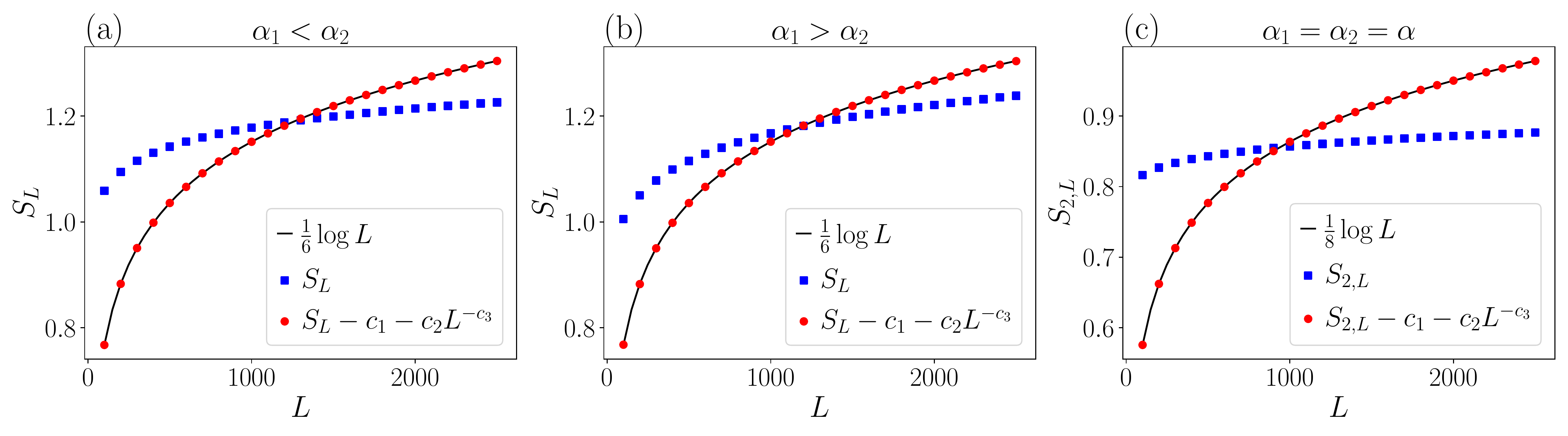}
	\caption{Numerical check of the entanglement scaling as a function of the subsystem size $L$ at the quantum critical point with chemical potential $h = -1+2^{1-\alpha_{1}}$ for different values of couplings power law decay exponents: $a)$ $\alpha_{1} = 1.5,\alpha_{2} = 1.8$, $b)$ $\alpha_{1} = 1.8,\alpha_{2} = 1.5$, $c)$ $\alpha_{1} = \alpha_{2} = 1.5$ and total system size $N = 2L$. As in Fig.\ref{fig: EntanglementScaling_hf}, blue squares represents the numerical data, the black solid line represents our analytical prediction for the scaling in the $L\gg 1$ limit, red dots are obtained from the numerics by subtracting the subleading corrections.}
	\label{fig: EntanglementScaling_ha}
\end{figure*}
This logarithmic scaling is analogous to the one obtained for a conformal field theory with central charge $c = 1/2$ \cite{Calabrese2009JPA}. This result is in agreement with previous findings \cite{Ares2015PRA,VodolaPRL2014} concerning the entanglement scaling in a Kitaev chain with long-range paring and nearest neighbors hopping $\alpha_1\to\infty$, here we show that the same scaling holds also for finite $\alpha_1$ as long as $\alpha_1>\alpha_2$. Figure \ref{fig: EntanglementScaling_hf}b shows the numerical check of the scaling behavior of the entanglement entropy $S_L = S_{1,L}$ for $\alpha_1>\alpha_2$ and $h = 1$. We obtain an excellent agreement once the subleading corrections are taken into account. In particular, we need to subtract from the numerical data the finite size corrections of the form
\begin{align}
	S_{L}-\frac{1}{6}\ln L = c_1+c_2L^{-c_3},\label{eq: subleading correctins}
\end{align}
where the $c_i = c_i(\alpha_1,\alpha_2,h)$, $i=1,2,3$, coefficients can be estimated from a fit with the numerical data. 

The most interesting case corresponds to the condition $\alpha_1 = \alpha_2 = \alpha$ which, as previously stated, is closely related to the long-range interacting quantum Ising chain. Moreover, we notice that in this regime the matrix symbol $G_k$, hosts non-commuting lateral limits as $k\to 0^\pm$ (see Eq.~\eqref{eq: lateral limits h1wlr}). This leads to the non-trivial dependence of the logarithmic contribution coefficient on $\alpha$
\begin{align}
	b_0(z) = \frac{2}{\pi^2}\left[\ln\frac{\sqrt{z^2-\sin^2(\alpha\pi/2)}+\cos(\alpha\pi/2)}{\sqrt{z^2-1}}\right]^2.
\end{align}
Inserting $b_0(z)$ in Eq.~\eqref{eq: Rényi integral} and performing the integration (see Appendix \ref{app: Coefficients of the Fisher-Hartwig expansion}), we obtain the logarithmic scaling behavior of the Rényi entropy
\begin{align}
	S_{\nu,L} = B_{\nu,\alpha}\ln L +\mathcal{O}(1), 
\end{align}
where
\begin{align}
	B_{\nu,\alpha} = \frac{1}{\pi^2(\nu-1)}\sum_{k=1}^\nu\arctan^2\left[\frac{\cos(\alpha\pi/2)}{\sqrt{\sin^2(\alpha\pi/2)+|z_{k,\nu}|^2}}\right],\label{eq: B alpha weak long range}
\end{align}
with $z_{k,\nu} = i\tan(\pi(2k-1)/2\nu)$. In particular, for $\nu = 2,3$, the sum in the previous expression reduces to
\begin{align}
	B_{2,\alpha} &= \frac{2}{\pi^2}\arctan^2\left[\frac{\cos(\alpha\pi/2)}{\sqrt{\sin^2(\alpha\pi/2)+1}}\right],
	\\
	B_{3,\alpha} &= \frac{1}{\pi^2}\arctan^2\left[\frac{\cos(\alpha\pi/2)}{\sqrt{\sin^2(\alpha\pi/2)+1/3}}\right].
\end{align}
\begin{figure}
	\centering
	\includegraphics[width=\linewidth]{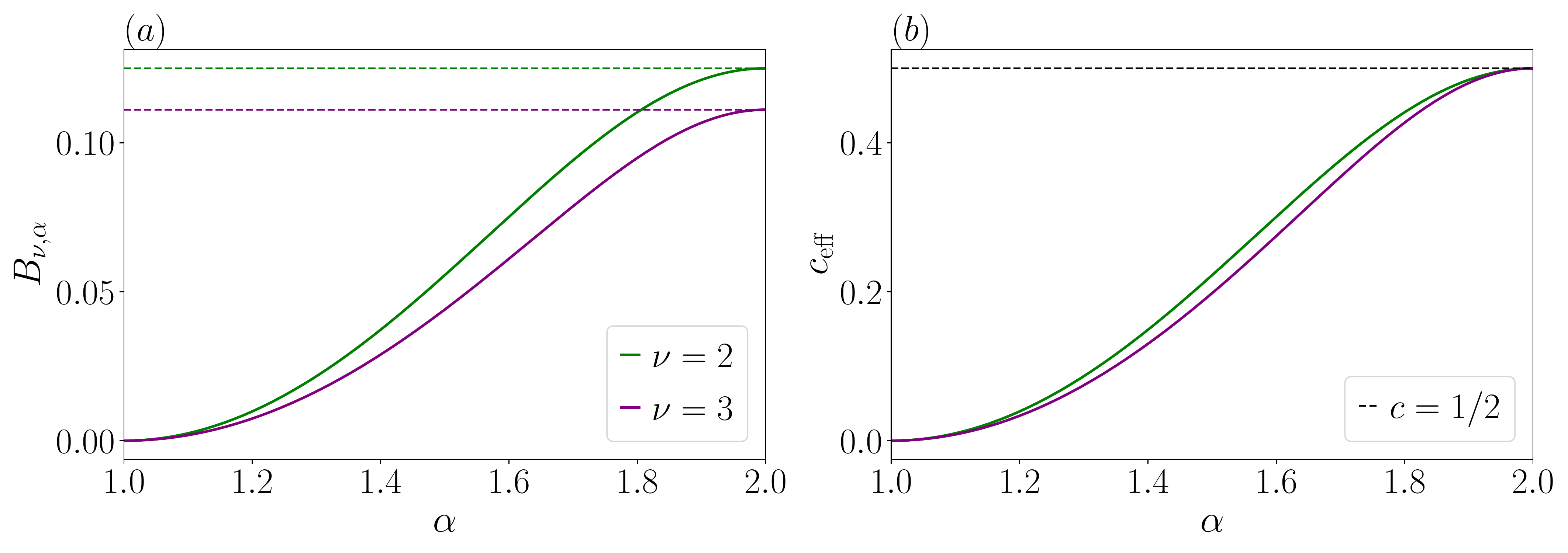}
	\caption{$a)$ Coefficient $B_{\nu,\alpha}$ of the logarithmic scaling of the $\nu$-Rényi entropy as a function of the power law decay exponent $\alpha =\alpha_1 = \alpha_2$, for $\nu = 2$ (green solid line) and $\nu = 3$ (purple solid line). The dashed lines correspond to the short-range values of the coefficients which are matched by the long-range ones for $\alpha = 2$. $b)$ effective central charge, obtained as $c_{\mathrm{eff}} = 6\nu B_{\nu,\alpha}/(\nu+1)$, as a function of $\alpha$ for $\nu = 2,3$. The black dashed line represents the central charge for nearest neighbor couplings $c = 1/2$.}
	\label{fig:B2a_weak_ceff}
\end{figure}
This analytical scaling of $S_{2,\nu}$ at $h = 1$ and for $\alpha_1=\alpha_2=\alpha$ is compared with the numerical result in Fig. \ref{fig: EntanglementScaling_hf}c. Also in this case, a good agreement is found once the subleading corrections~\eqref{eq: subleading correctins} are taken into account. 

We note that the expression for the scaling coefficients in Eq.\eqref{eq: B alpha weak long range} is valid only for integers $\nu>1$. Indeed, in this case $ds_{\nu}/dz$ is a meromorphic function with poles located on the imaginary axis. This allows us to evaluate the integral in \eqref{eq: Rényi integral} by summing over the residues at these poles  (see Appendix \ref{app: Coefficients of the Fisher-Hartwig expansion} for details on the calculation). On the other hand, for $\nu = 1$, we have that
\begin{align}
	\frac{ds_{\nu = 1}(1+\epsilon,z)}{dz} = \ln\left(\frac{1+\epsilon-z}{1+\epsilon+z}\right),
\end{align}
which has two branch cuts from $\pm(1+\epsilon)$ to infinity (see Appendix \ref{app: Coefficients of the Fisher-Hartwig expansion}). Therefore, to evaluate the integral in Eq.~\eqref{eq: Rényi integral} for $\nu = 1$, we perform the integration along these cuts and take into account the change in the phase of the logarithm when we go around the branch points. This reduces the integral to two real integrals, which we evaluate numerically. In the case where $\alpha_{1} = \alpha_{2} = \alpha$ and $h = 1$, the integrand still depends on $\alpha$ even for $\nu = 1$, so we can still expect the coefficient for the logarithmic divergence of the von Neumann entropy $S_{1,L}$ to have a nontrivial $\alpha$ dependence.

It is important to observe that at variance with the $\alpha_1\neq \alpha_{2}$ cases, the scaling coefficient $B_{\nu,\alpha}$ cannot be written in the form
\begin{align}
\label{cft_const}
	B_{\nu,\alpha}\neq B_{\nu,\mathrm{CFT}} = \frac{\nu+1}{6\nu}c,
\end{align}
where $c$ is the central charge of some conformal field theory describing the model at the quantum critical point. This observation supports our previous claim that the case $\alpha_{1}=\alpha_{2}$ is special and, somehow, closer to the one of a strongly interacting system such as the long-range Ising model. Indeed, while the case $\alpha_1\neq \alpha_{2}$ continues to obey the r.h.s. of Eq.\,\eqref{cft_const} and, so, is more likely to be described by a CFT, the case $1<\alpha_1 = \alpha_2<2$ goes beyond this description as the scaling of the ground state entanglement at the critical point cannot be related to the universal properties of a conformal field theory. A similar result is expected for the  Ising model in a transverse field, where the inclusion of long-range interactions is expected to increase the effective dimension of the model and, so, disrupt any CFT description.

 Figure \ref{fig:B2a_weak_ceff}a shows the coefficients $B_{\nu,\alpha}$ for $\nu = 2,3$ as a function of $\alpha\in[1,2]$, we notice that the value of the logarithmic scaling coefficients starts from zero at $\alpha = 1$ and then grows with $\alpha$ reaching the short-range value for $\alpha = 2$. Moreover, Fig.\,\ref{fig:B2a_weak_ceff}b shows the $\alpha$ dependence of the effective central charge defined as $c_\mathrm{eff}(\alpha) = 6\nu B_{\nu,\alpha}/(\nu+1)$ as a function of $\alpha$. We notice that, apart from the extrema $c_\mathrm{eff}(1) = 0$ and $c_\mathrm{eff}(2) = 1/2$, the effective charge also depends on the Rényi entropy order $\nu$, thus confirming the fact that it cannot be considered as the proper central charge of a conformal field theory. These results are in agreement with the findings of Ref.\,\cite{Lepori2016AnnPhys}, where the breakdown of conformal symmetry in a long-range fermionic chain was established.

Finally, we consider the non-homogeneous critical point $h = -1+2^{1-\alpha_1}$. In this case, the power of the dispersion relation near the soft mode $k = \pi$ is not affected by the presence of long-range couplings (see Appendix \ref{app: Dispersion relation around the critical modes}). Accordingly, also the symbol discontinuity is independent of the value of $\alpha_{1,2}$, in particular, we find
\begin{align}
	G_\pi^\pm = \lim_{k\to \pi^\pm}G_k =\pm\sigma_y,\quad \forall \alpha_{1},\alpha_{2}>1.
\end{align}
This leads to a logarithmic contribution coefficient
\begin{align}
	b_{\pi}(z) =  \frac{1}{2\pi^2}\left(\ln\left(\frac{z+1}{z-1}\right)\right)^2.
\end{align}
The corresponding scaling of the entanglement entropy is then the one obtained in Eq.~\eqref{eq: entanglement scaling c= 1/2}, which is equivalent to the entanglement scaling in the nearest neighbor Kitaev chain, at a quantum critical point characterized by a conformal field theory with central charge $c = 1/2$. Figure \ref{fig: EntanglementScaling_ha} shows the entanglement scaling behavior at the non-homogeneous critical point $h = -1+2^{1-\alpha_1}$ with $\alpha_1<\alpha_2$ (Fig. \ref{fig: EntanglementScaling_ha}a), $\alpha_1>\alpha_2$ (Fig. \ref{fig: EntanglementScaling_ha}b) and $\alpha_1=\alpha_2$ (Fig. \ref{fig: EntanglementScaling_ha}c). Also in this case a nice agreement with the theoretical prediction in the thermodynamic limit is found once finite size corrections are taken into account.

The results for the entanglement scaling with the subsystem size in at different critical points and for different values of the $\alpha_{1},\alpha_{2}$ parameters within the weak long-range regime considered in this section ($1<\alpha_{1},\alpha_{2}<2$) are summarized in Table \ref{table: weak long-range}.
\begin{table}[h!]
\centering
\renewcommand{\arraystretch}{1.5}
\begin{tabular}{|c|c|c|c|}\hline
	 &$1<\alpha_1<\alpha_{2}<2$ &$1<\alpha_2<\alpha_{1}<2$ &$1<\alpha_2=\alpha_{1}<2$\\
	\hline
	$h = 1$  &$S_{\nu,L} = \mathcal{O}(1)$ &$S_{\nu,L} \approx\frac{\nu+1}{12\nu}\ln L$ &$S_{\nu,L} \approx B_{\nu,\alpha}\ln L$\\
	\hline
	$h = -1+2^{1-\alpha_1}$ &$S_{\nu,L}\approx \frac{\nu+1}{12\nu}\ln L$ &$S_{\nu,L}\approx \frac{\nu+1}{12\nu}\ln L$ &$S_{\nu,L}\approx \frac{\nu+1}{12\nu}\ln L$\\
	\hline
\end{tabular}
\caption{Summary of entanglement scaling results at different quantum critical points and for various values of $\alpha_{1}$ and $\alpha_{2}$ in the weak long-range regime. The symbol $\approx$ denotes equality up to subleading $\mathcal{O}(1)$ corrections.}
\label{table: weak long-range}
\end{table}
\section{Strong long-range regime}\label{sec: Strong long-range regime}
The situation in the strong long-range regime is more involved. In particular previous studies on fermionic systems with strong long-range pairing interactions \cite{VodolaPRL2014,Ares2015PRA,Ares2018PRA} reported logarithmic violations of the entanglement area law even away from criticality. However, in those cases, the noncritical logarithmic scaling of the ground state entanglement was associated with divergences in the long-range couplings due to the fact that no Kac scaling was introduced in the model Hamiltonian. Therefore, one may think such anomalous scalings to be trivially related to the loss of the system extensivity. On the other hand, as shown in Sec. \ref{sec: Kitaev chain with long-range couplings}, the introduction of a Kac scaling in the Hamiltonian allows us to define a model with strong long-range interaction still preserving the energy extensivity.

In particular, when a Kac scaling is introduced, the coupling divergences for $\alpha_1,\alpha_2<1$ are canceled, and accordingly also the symbol discontinuity associated with them disappears. However, an infinite number of new nontrivial discontinuities arise due to the fact that the spectrum becomes discrete also in the thermodynamic limit. More precisely, as a consequence of the spectrum discontinuity, the symbol becomes discontinuous for any $k = 2\pi n/N$. Indeed, in the thermodynamic limit, $G_k$ reads
 %Indeed the proper definition of limit $k = 2\pi n/N\to q^\pm =2\pi m/N, 2\pi(m+ 1)/N$ reads
%\begin{align}
	%\lim_{k\to q^\pm} G_{k} = \lim_{N\to\infty}\lim_{k\to q,q+\frac{2\pi}{N}}G_k = \lim_{n\to m, m+1}G_n = G_{m,m+1},
%\end{align}
%
%where we have introduced  
%
\begin{align}
\lim_{N\to\infty}G_k = G_n = \frac{2(h-\tilde{t}_n)}{\omega_n}\sigma_z-\frac{2\tilde{\Delta}_n}{\omega_n}\sigma_y.\label{eq: Gn}
\end{align}
Then it can be labeled by a discrete integer number $n$, while the $k$ variable becomes continuous. More precisely, any real physical implementation of the model has necessarily a finite size. Therefore, the actual physical meaning of the continuum limit as $N\to \infty$ is that the difference between two consecutive values of $k$ is of order $\mathcal{O}(N^{-1})$. However, in the strong long-range case, a difference of order $\mathcal{O}(N^{-1})$ in the $k$ variable results in a finite jump of the spectrum $\omega_n$ which remains discrete even in the thermodynamic limits, thus resulting in a discontinuity of the matrix symbol $G_k$ for any $k$ independently of the value of the chemical potential $h$. In particular, since for any $\alpha_{1,2}>0$ or $\alpha_{1} = \alpha_{2} = 0$ and $h \neq 0$ the many-body ground state is still the Bogoliubov vacuum, then the two lateral limits corresponding to a given $k^\pm = 2\pi n/N, 2\pi(n+1)/N$ can be written as  
\begin{align}
	G_k^{\pm} = \begin{cases}
		G_{n+1} = \cos\phi_{n+1}\sigma_z +\sin\phi_{n+1}\sigma_y\\
		G_{n} = \cos\phi_n\sigma_z +\sin\phi_n\sigma_y
	\end{cases},
\end{align}
where we have introduced the angles $\phi_n$ defined by the conditions $\cos\phi_n = 2(h-\tilde{t}_n)/\omega_{n}$ and $\sin\phi_n = -2\tilde{\Delta}_n/\omega_{n}$. Then, following the analytic procedure introduced in Section \ref{sec: Entanglement scaling}, for any value of $h$, we obtain a logarithmic scaling of the ground state Rényi entropies of the form
\begin{align}
	S_{\nu,L} = B_{\nu}(h)\ln L +\mathcal{O}(1),
\end{align}
where the  $B_{\nu}(h)$ coefficient is a function of $\nu ,\alpha_1,\alpha_2$ and $h$. Then $B_{\nu,\alpha_1,\alpha_2}(h)$ is given by the sum of $N$ contributions corresponding to the $N$ discontinuities of the symbol, reading
\begin{align}
	B_{\nu}(h) = \sum_{n = -N/2+1}^{N/2} B_{\nu}^{(n)}(h), 
\end{align} 
where, as shown in Appendix \ref{app: Coefficients of the Fisher-Hartwig expansion}, each contribution reads 
\begin{align}
	B_{\nu}^{(n)}(h) = \frac{1}{\pi^2(\nu-1)}\sum_{l = 1}^\nu\arctan^2\left[\frac{\sin((\phi_{n+1}-\phi_n)/2)}{\sqrt{\cos^2((\phi_{n+1}-\phi_n)/2)+|z_l|^2}}\right].\label{eq: B strong long-range}%\\
	%&=\frac{1}{\pi^2(\nu-1)}12\sum_{k = 1}^{\nu}\arctan^2\sqrt{\frac{\omega_{n+1}\omega_{n-1}-(h-t_{n+1})(h-t_{n-1})-\Delta_{n+1}\Delta_{n-1}}{(2|z_k|^2+1)\omega_{n+1}\omega_{n-1}+(h-t_{n+1})(h-t_{n-1})+\Delta_{n+1}\Delta_{n-1}}}
\end{align}
where $|z_l|^2 = \tan^2(\pi(2l-1)/2\nu)$, with $l = 1,\dots,\nu$ and $l\neq (1+\nu)/2$. In particular, for $\nu = 2$ the above sum can be written explicitly as
\begin{align}
	B_{2}^{(n)}(h) = \frac{2}{\pi^2}\left[\arctan\sqrt{\frac{\omega_{n+1}\omega_{n}-(h-\tilde{t}_{n+1})(h-\tilde{t}_n)-\tilde{\Delta}_{n+1}\tilde{\Delta}_n}{3\omega_{n+1}\omega_{n}+(h-\tilde{t}_{n+1})(h-\tilde{t}_n)+\tilde{\Delta}_{n+1}\tilde{\Delta}_n}}\right]^2.\label{eq: B2n}
\end{align}

As we have already seen in the previous Sections, the most interesting situation is the one with equally long-range hopping and pairing amplitudes, i.e., with $\alpha_1 = \alpha_{2} = \alpha$, while we expect only minor differences to appear when $\alpha_{1}\neq \alpha_{2}$, as long as they are both smaller than the system dimension (here d = 1). Therefore, for the sake of simplicity, we will limit our treatment to the $\alpha_{1} = \alpha_{2} = \alpha$ case in the following analysis of the strong long-range regime.
\begin{figure*}
	\centering
	\includegraphics[width=\linewidth]{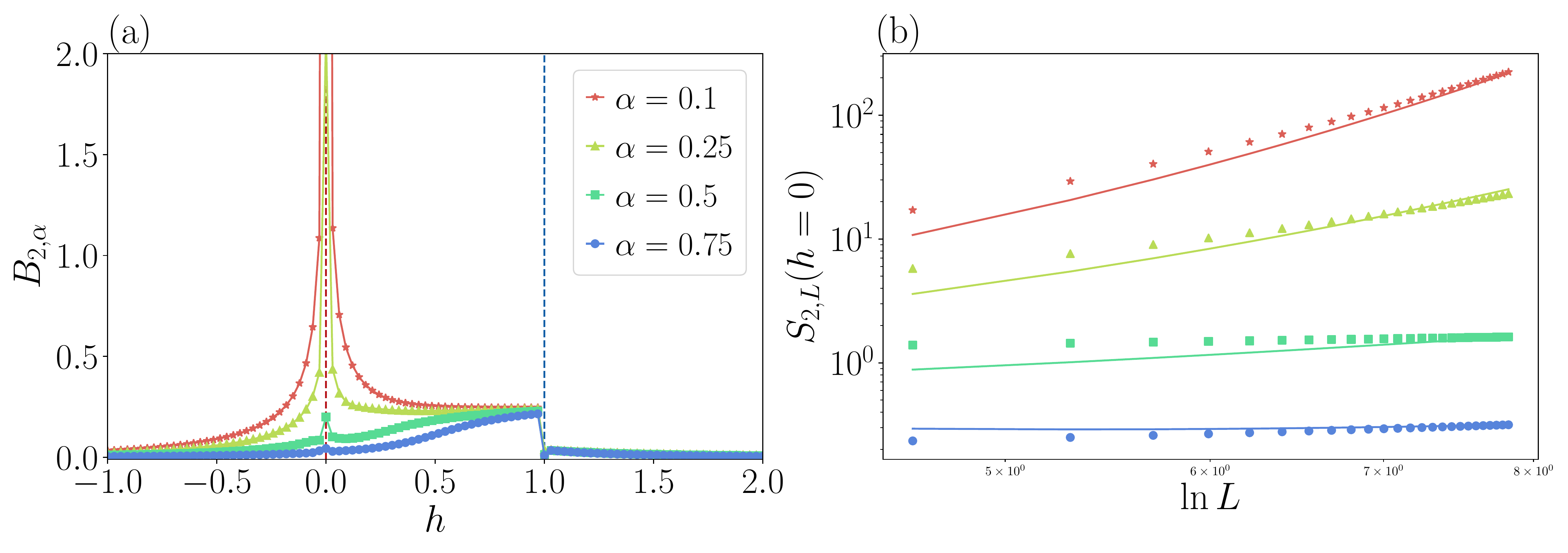}
	\caption{a) Rényi-2 scaling coefficient $B_{2,\alpha}$ as a function of the chemical potential $h$ for different values of the power law decay coefficient $0<\alpha = \alpha_1 = \alpha_{2}<1$. The red and blue vertical lines correspond to the $h = 1$ and $h = 0$ critical points, respectively. b) Numerical check for the entanglement subvolume law scaling at $h = 0$ for different values of $0<\alpha<1$, plotted as a function of the logarithm of the subsystem size $\ln L$. The total system size is taken to be $N = 2L$. Scattered points represent the numerical data for the half-chain Rényi-2 entropy, while the solid lines represent our prediction $B_{2,\alpha}\ln L$.}
	\label{fig: StrongLR}
\end{figure*}

Figure \ref{fig: StrongLR}a shows $B_{2}(h)$ as a function of the chemical potential $h$ for different values of $\alpha_{1} = \alpha_{2} = \alpha$. First of all, we notice that for any values of the chemical potential $h\neq 0$ and of $\alpha>0$ the scaling coefficient is of order $B_{2}(h\neq 0) = \mathcal{O}(1)$, then leading to a logarithmic violation of the area law even away from the quantum critical points. Moreover, two singularities appear at the quantum critical points $h = \tilde{t}_k,\tilde{t}_\pi = 1,0$. In particular, we have a discontinuity for $h = 1$ and a divergence with the subsystem size for $h = 0$, leading to a subvolume law entanglement scaling. 

These facts can be understood as follows. The spectrum is labeled by the discrete index $n$ leading to a finite gap between the ground state and the first excited levels which are associated with discontinuities of the symbol. However, for $n\gg 1$ all the modes accumulate around $\omega_\infty = 2|h|$. This means that an extensive number of single-particle states is almost degenerate. Consequently, as long as $h\neq 0$, we may expect only the first few modes around $n = 0$ to provide a significant contribution to the symbol discontinuity leading to a coefficient $B_\nu(h\neq 0) =\mathcal{O}(1)$. Accordingly, we may expect many features of the entanglement scaling coefficients for values of the chemical potential sufficiently far from the $h = 0$ point, to be qualitatively reproduced by considering a single discontinuity approximation in which only the first discontinuity between the $n = 0$ and the first two degenerate levels $n = \pm 1$ is considered, i.e., $B_\nu(h\neq 0)\approx B_\nu^{(0)}+B_\nu^{(-1)}$. Then, as detailed in Appendix \ref{app: Discontunities in the strong long-range regime} within this approximation the discontinuity coefficient reads
\begin{align}
	B_\nu(h\neq 0) \approx 
	\frac{2}{\pi^2(\nu-1)}\sum_{l = 1}^\nu\arctan^2\left[\frac{\cos(\phi_1/2)}{|z_l|^2+\sin^2(\phi_1/2)}\right]  \quad&\mathrm{if}\quad h<1,\\
	B_\nu(h\neq 0) \approx 	\frac{2}{\pi^2(\nu-1)}\sum_{l = 1}^\nu\arctan^2\left[\frac{\sin(\phi_1/2)}{|z_l|^2+\cos^2(\phi_1/2)}\right]  \quad&\mathrm{if}\quad h>1.
\end{align}
This approximation then allows us to capture the origin of the scaling coefficient  discontinuity at $h = 1$. This originates from the fact that the zero mode gives different contributions at the two sides of the transition, indeed (see Appendix \ref{app: Discontunities in the strong long-range regime})
\begin{align}
	\phi_0 = \arccos[\mathrm{sign}(h-1)] = \begin{cases}
		\pi &\mathrm{if}\quad h<1\\
		0   &\mathrm{if}\quad h>1
	\end{cases}.
\end{align} 

The single discontinuity approximation turns out to correctly reproduce the qualitative features as long as the chemical potential $h$ is sufficiently far from $h = 0$ and for sufficiently large power law decay exponent $\alpha>1/2$. On the other hand, this simple approximation is no more accurate as the chemical potential approaches the $h = 0$ point. Indeed, in the zero chemical potential case $\omega_\infty = 0$, and more precisely $\omega_n$, $\tilde{t}_n$ and $\tilde{\Delta}_n$ approach their asymptotic values differently if we consider the even or the odd modes (see Appendix \ref{app: Discontunities in the strong long-range regime} for more details). As a consequence, for sufficiently small $\alpha$, the number of relevant symbol discontinuities grows as a power law of the subsystem size $L$, leading to a fractal subvolume-law entanglement scaling. In particular, using the asymptotic expansion of $\omega_n$, $\tilde{t}_n$ and $\tilde{\Delta}_n$ in the $n\to\infty$ limit we can extract the leading order dependence of $B_\nu(h=0)$ from $L$, which, as shown in Appendix \ref{app: Discontunities in the strong long-range regime}, reads 
\begin{align}
	B_\nu(h=0) = \begin{cases}
		\mathcal{O}(L^{1-2\alpha}) &\mathrm{if}\quad \alpha<1/2\\
		\mathcal{O}(1) &\mathrm{if}\quad \alpha>1/2
	\end{cases}.
\end{align}
Accordingly, the leading order contribution to the entanglement Rényi entropy of the system ground state at zero chemical potential takes the nontrivial form
\begin{align}
	S_{\nu,L}(h=0) = \begin{cases}
		\mathcal{O}(L^{1-2\alpha}\ln L)&\mathrm{if}\quad \alpha<1/2\\
		\mathcal{O}(\ln L) &\mathrm{if}\quad \alpha>1/2
	\end{cases}.\label{eq: entanglement scaling strong LR h0}
\end{align}
This analytic result matches the numerics in the large $L$ limit. This is shown in Fig.\,\ref{fig: StrongLR}b, where the numerical and analytical results for $S_{2,L}$ are plotted as a function of $\ln L$ and for different values of $\alpha$. It is important to notice that approaching the thermodynamic limit in the $h = 0$ case the spectrum becomes increasingly more degenerate approaching the $\alpha = 0$ case. Then, for each finite $N$, a large number of states nearly degenerate with the ground state exists, making the estimate of the subleading corrections scaling technically challenging.

Finally, as already stated in Sections \ref{sec: Kitaev chain with long-range couplings} and \ref{sec: Entanglement scaling}, the mean-field case with $\alpha_{1} = \alpha_{2} = 0$ and $h = 0$ must be treated separately. Indeed, in this case the ground state degeneracy allows for a finite fermionic population of the even Bogoliubov modes, $f_n\neq 0$ $\forall n (\mathrm{even})$, this leads to the entanglement scaling 
\begin{align}
	S_{\nu,L}(\alpha = 0,h = 0) = \frac{1}{1-\nu}\sum_{n(\mathrm{even})}\ln\left[(1-f_n)^\nu+f_n^\nu\right] +\mathcal{O}(\ln L).
\end{align}
In particular the maximal Rényi entropy is reached when $f_n = 1/2$ $\forall n (\mathrm{even})$
\begin{align}
	S_{\nu,L}^{\mathrm{max}}(\alpha = 0,h = 0) = N_0\ln 2 +\mathcal{O}(1)= \frac{L}{2}\ln 2  +\mathcal{O}(1),
\end{align}
where $N_0$ is the number of zero modes, which in this case corresponds to the number of even modes $N_0\simeq L/2$ and the subleading corrections are at most of order $\mathcal{O}(1)$. Indeed, as shown in Appendix \ref{app: Coefficients of the Fisher-Hartwig expansion}, the discontinuity coefficients $B_\nu$ which would lead to logarithmic corrections turn out to be exactly zero when all the even fermionic populations are $f_{n(\mathrm{even})} = 1/2$. Moreover, we notice that the maximal Rényi entropy that we have obtained employing the Fisher-Hartwig expansion corresponds to the largest possible entropy allowed by the ground state degeneracy
\begin{align}
	S_{\nu,L}^{\mathrm{max}}(\alpha = 0,h = 0) = \ln \mathrm{Deg}[|\mathrm{gs}_{\alpha = 0,h = 0}\rangle] = N_0\ln 2.
\end{align}
This tells us that the Fisher-Hartwig result, obtained as a large subsystem size expansion, actually becomes exact in this maximally entangled case. 

The results for the entanglement scaling with the subsystem size for different values of the $h$ and $\alpha = \alpha_{1}=\alpha_{2}$ parameters within the strong long-range regime considered in this section ($0<\alpha<1$) are summarized in Table \ref{table: strong long-range}.
\begin{table}[h!]
	\centering
	\renewcommand{\arraystretch}{1.5}
	\begin{tabular}{|c|c|c|c|}\hline
		&$\alpha = 0$ &$0<\alpha<1/2$ &$1/2<\alpha<1$\\
		\hline
		$h \neq 0$  &$S_{\nu,L} = \mathcal{O}(\ln L )$ &$S_{\nu,L} = \mathcal{O}(\ln L )$ &$S_{\nu,L} = \mathcal{O}(\ln L )$\\
		\hline
		$h = 0$ &$S_{\nu,L} = \mathcal{O}(L)$ &$S_{\nu,L} = \mathcal{O}(L^{1-2\alpha}\ln L)$ &$S_{\nu,L}=\mathcal{O}(\ln L)$\\
		\hline
	\end{tabular}
	\caption{Summary of entanglement scaling results at different quantum critical points and for various values of $\alpha = \alpha_{1} = \alpha_{2}$ in the strong long-range regime.}
	\label{table: strong long-range}
\end{table}
\section{Conclusion and outlooks}
In this paper, we have further extended the understanding of the peculiar properties of entanglement in quantum systems featuring long-range interactions. At this scope, we have investigated, as a paradigmatic example, the ground state entanglement scaling of a spinless fermionic chain with long-range hopping and pairing amplitudes. The simplicity of the model and its truly non-additive nature allowed us to unveil an extremely rich and non-trivial phenomenology, which we have fully characterized both numerically and analytically in the different regions of the relevant parameters, i.e., the power law decay exponents of the hopping and pairing couplings $\alpha_{1}$, $\alpha_{2}$ and the chemical potential $h$. In particular, two main regimes may be distinguished: the weak long-range regime with $1<\alpha_1,\alpha_{2}<2$ and the strong long-range regime with $0<\alpha_1,\alpha_{2}<1$. 

In the weak long-range case, the system quasiparticle spectrum becomes continuous in the thermodynamic limit and the main effect of the non-local couplings is to change the dispersion relation near the gapless critical modes. Accordingly, the standard area law, typical of gapped local Hamiltonians, is satisfied in this regime apart from the logarithmic violations which appear in correspondence of the two quantum critical points located at $h = 1,-1+2^{1-\alpha_1}$. Such logarithmic scaling of the ground state Rényi entropies is related to discontinuities in the symbol of the correlation matrix which is a block Toeplitz matrix. The fact that the contribution to the entanglement scaling of each discontinuity only depends on the value of the symbol \cite{Ares2015PRA,Ares2018PRA} at each side of the jump, allowed us to exactly compute its coefficients. Most significantly, when the hopping and pairing couplings are equally long-range, i.e., $\alpha_{1} = \alpha_{2} = \alpha$, the coefficient in front of the critical logarithmic divergence at $h = 1$ turns out to have a non-trivial dependence on $\alpha$~\eqref{eq: B alpha weak long range}. 

Interestingly, the coefficient $B_{\nu,\alpha}$ is of non-universal nature, since it originates from the precise form of the spectrum in the proximity of the critical modes, and not only from the dispersion relation power law exponent. As a consequence, the critical entanglement scaling is not compatible with the result obtained from any conformal field theory and our result may be seen as a benchmark of the fact that the presence of long-range couplings explicitly breaks the critical conformal symmetry\,\cite{Lepori2016AnnPhys}. These findings  demonstrate the peculiarity of the $\alpha_{1}=\alpha_{2}$ case, whose physics is expected to be, and indeed is, closer to the one of a strongly interacting system such as the quantum Ising model, where long-range couplings are expected to increase the effective dimension and, so, disrupt integrability\,\cite{defenu2017criticality}.

Moreover, for $\alpha_{1}\neq\alpha_{2}$, the critical entanglement scaling becomes $\alpha$ independent. In particular, when  $\alpha_{1}>\alpha_{2}$, i.e., the pairing coupling has a slower decay with respect to the hopping, the entanglement scaling is compatible with that of conformal field theory with central charge $c = 1/2$. This is in agreement with the results of Ref.\,\cite{Ares2015PRA,Ares2018PRA}, where a Kitaev chain with long-range pairing and nearest neighbors hopping is considered, the validity of such results is then here extended to any long-range hopping with power law decay exponent $\alpha_{1}>\alpha_{2}$. The strong anisotropy between the case of dominating hopping $\alpha_{1}<\alpha_{2}$ and the case of dominating paring $\alpha_{1}>\alpha_{2}$ is typical of the long-range Kitaev chain\,\cite{DefenuPRB2019}.

In the strong long-range regime, the situation is more involved, indeed the quasiparticle spectrum can no more be considered continuous in the thermodynamic limit. Consequently, the matrix symbol of the block Toeplitz correlation matrix formally becomes discontinuous at every point of the spectrum. However, as shown in Section \ref{sec: Strong long-range regime}, in most situations only a few of such discontinuities truly contribute to the entanglement scaling, leading to a logarithmic dependence on the subsystem size even outside criticality. Also in this case the coefficients of such logarithmic divergence can be computed analytically for different values of the parameters $\alpha_{1,2}$ and $h$. 

The most interesting situation turns out to be the zero chemical potential point $h = 0$ in the strong long-range regime. Indeed, in this case, the coefficient in front of the critical logarithmic entanglement scaling diverges as a power law of the subsystem size, leading to a fractal subvolume-law entanglement scaling. More precisely, we were able to analytically extract the leading entanglement dependence from the subsystem size, which turns out to be of the form $S_{\nu,L}\approx L^{1-2\alpha}\ln L$, with $0<\alpha = \alpha_1 = \alpha_{2}<1/2$, where $S_{\nu,L}$ is any $\nu$-Renyi entropy with $\nu>1$. Similar sub-volume laws have already been observed in different (more complex) scenarios and, in particular, in the entanglement scaling of measurement induced phase transitions\,\cite{Block2022PRL}, where they arise due to the suppression of entanglement caused by repeated measurements in a long-range systems. Here, this phase emerges naturally in the equilibrium scaling, but it needs stronger interactions to appear with respect to the dynamical case.

Finally, in the completely mean-field case, the system presents an extensive number of degenerate modes with zero energy. These zero modes can be populated also in the many-body ground state whose degeneracy then grows exponentially with the number of zero modes. Consequently, the ground state entanglement shows a volume law behavior proportional to the size of the considered subsystem $S_{\nu,L}(\alpha = 0)\approx L$. 

Our studies evidence that long-range couplings can greatly improve the scaling of entanglement at equilibrium and, therefore, that long-range interacting quantum systems represent the ideal candidate for reliable and robust quantum computation. Nevertheless, such fostered entanglement properties may not persist out-of-equilibrium, since long-range interactions have  been shown to suppress the dynamical spread of entanglement in certain systems\,\cite{Lerose2020PRR}. For the future, we intend to investigate these issues by performing quantum simulations of the model on actual quantum computers. This demands a careful engineering of the artificial non-local couplings on local quantum devices, a task which we are currently tackling on IBM Quantum devices\,\cite{SolfanelliIBMQ}.
%In the future, we intend to investigate these issues by performing a quantum simulation of the model on a real quantum computer. Performing a similar quantum computation on current publicly available hardwares necessitates to improve our understanding of how to engineer artificial non-local couplings on local quantum devices, a task which we are currently tackling\,\cite{SolfanelliIBMQ}.

The rich phenomenology hosted by the minimal long-range model we considered, already at equilibrium, suggests that many of the intriguing dynamical phenomena which are recently emerging in the quantum community, such as the non-trivial fractal entanglement scalings in the contest of measurement-induced entanglement transitions\,\cite{xuPhysics2022}, can be simply ascribed to the presence of sufficiently long-range couplings among the microscopic components of the model, without any need of further complexity in the physical system under consideration. Further work is needed in order to investigate the dynamical properties of entanglement in the Kitaev chain with long-range pairing and hopping couplings subjected to a unitary or a non-unitary (measurement-like) evolution. These interesting problems are beyond the scope of this work and we leave them as an outlook for future projects.
\subsection*{Acknowledgements}
We acknowledge support by the Deutsche Forschungsgemeinschaft (DFG, German Research Foundation)   under Germany’s Excellence Strategy 
EXC2181/1-390900948 (the Heidelberg STRUCTURES Excellence Cluster). This work is part of the MIUR-PRIN2017 project Coarse-grained description for nonequilibrium systems and transport phenomena (CO-NEST) No. 201798CZL. AS and SS acknowledge acknowledge financial support from National Centre for HPC, Big Data and Quantum Computing (CN00000013). Access to the IBM Quantum Computers was obtained through 
the IBM Quantum Hub at CERN.

\newpage
\appendix
%
%%%%
\section{Derivation of the matrix symbol}\label{app: Derivation of the matrix symbol}
In this Appendix we provide the details for the derivation of the matrix symbol in Eq.~ \eqref{eq: matrix symbol} of the main text. We start from the definition of the correlation matrix of a stationary state $|\psi\rangle$, then passing to the Fourier basis we obtain
\begin{align}
	G_k = 2\langle\psi|\begin{pmatrix}
		\hat{c}_k\\
		\hat{c}^\dagger_{-k}
	\end{pmatrix}\begin{pmatrix}
		\hat{c}^\dagger_k &\hat{c}_{-k}
	\end{pmatrix} |\psi\rangle -I. 
\end{align}
Introducing the Bogoliubov transformation 
\begin{align}
	\begin{pmatrix}
		\hat{\gamma}_k\\
		\hat{\gamma}^\dagger_{-k}
	\end{pmatrix} =  U_k \begin{pmatrix}
		\hat{c}_k\\
		\hat{c}^\dagger_{-k}
	\end{pmatrix},\quad U_k = \begin{pmatrix}
		\cos\theta_k/2   &i\sin\theta_k/2\\
		-i\sin\theta_k/2 &-\cos\theta_k/2
	\end{pmatrix},
\end{align}
we can write the symbol in terms of the Bogoliubov modes as
\begin{align}
	G_k = 2U_k^\dagger\langle\psi|\begin{pmatrix}
		\hat{\gamma}_k\\
		\hat{\gamma}^\dagger_{-k}
	\end{pmatrix}\begin{pmatrix}
		\hat{\gamma}^\dagger_k &\hat{\gamma}_{-k}
	\end{pmatrix} |\psi\rangle U_k -I. \label{eq: Gk}
\end{align}
We now compute the expectation value in a stationary state associated to the fermionic populations of the Bogoliubov modes $f_k = \langle \hat{\gamma}_k^\dagger \hat{\gamma}_k\rangle$, so that
\begin{align}
	2\langle\psi|\begin{pmatrix}
		\hat{\gamma}_k\\
		\hat{\gamma}^\dagger_{-k}
	\end{pmatrix}\begin{pmatrix}
		\hat{\gamma}^\dagger_k &\hat{\gamma}_{-k}
	\end{pmatrix}|\psi\rangle -I= 
	\begin{pmatrix}
		1-2f_k &0\\
		0      &2f_{k}-1
	\end{pmatrix}.
\end{align}
Finally, inserting this expectation value in Eq.\,\eqref{eq: Gk} and using the definition of the Bogoliubov angles $\tan\theta_k = \tilde{\Delta}_k/(h-\tilde{t}_k)$ we obtain
\begin{align}
	G_k = 2(1-(f_k+f_k))\left[\frac{h-\tilde{t}_k}{\omega_k}\sigma_z-\frac{\tilde{\Delta}_k}{\omega_k}\sigma_y\right]-(f_k-f_{-k})I,
\end{align}
which is the expression for the matrix symbol used in the main text. 

\section{Coefficients of the Fisher-Hartwig expansion}\label{app: Coefficients of the Fisher-Hartwig expansion}

The general form of the matrix symbol in Eq.~\eqref{eq: Gk} can be used to compute the different terms in the Fisher-Hartwig expansion of the Rényi entropies for large subsystem size in every situation considered in the main text. For this purpose, it is useful to rewrite $G_k$ as
\begin{align}
	G_k = a_k\left[\cos\phi_k\sigma_z+\sin\phi_{k}\sigma_y\right]+b_kI, 
\end{align}
where we have introduced the coefficients $a_k = 1-(f_k+f_{-k})$ and $b_k =f_{-k}-f_k$ and the angle $\phi_k$ such that $\cos\phi_k = 2(h-\tilde{t}_k)/\omega_k$ and $\sin\phi_{k} = -2\tilde{\Delta}_k/\omega_k$. 

Let us start from the first term of the expansion in Eq.~\eqref{eq: Fisher-Hartwig expansion} this is obtained by first computing the determinant 
\begin{align}
	\det\left[z\mathbb{I}-G_k\right] = (z-b_k)^2-a_k^2,
\end{align}
Then, the contribution to first term in the entanglement scaling coming from each $k$-mode is obtained from the integral
\begin{align}
	\mathcal{S}_k %&= \lim_{\epsilon\to 0^+}\oint_{\mathcal{C}}\frac{dz}{4\pi i} s_\nu(1+\epsilon,z)\frac{d\ln\left[z\mathbb{I}-G_k\right]}{dz}
	%\\\notag
	&= \lim_{\epsilon\to 0^+}\oint_{\mathcal{C}}\frac{dz}{2\pi i} s_\nu(1+\epsilon,z)\frac{(z-b_k)}{(z-b_k)^2-a_k^2}
	\\\notag
	&=\frac{1}{2}\left[s_\nu(1,b_k+a_k)+s_\nu(1,b_k-a_k)\right]
	\\\notag
	&=\frac{1}{2(1-\nu)}\left[\ln(f_k^\nu+(1-f_k)^\nu)+\ln(f_{-k}^\nu+(1-f_{-k})^\nu)\right],
\end{align}
where Cauchy’s residue theorem and the expression~\eqref{eq: s_nu} for $s_\nu(x,y)$ have been used. Finally, summing over all the modes and using the $k\to-k$ symmetry we obtain
\begin{align}
	\sum_k \mathcal{S}_k = \frac{1}{1-\nu}\sum_k \ln(f_k^\nu+(1-f_k)^\nu).
\end{align}
\begin{figure}
	\centering
	\includegraphics[width=0.8\linewidth]{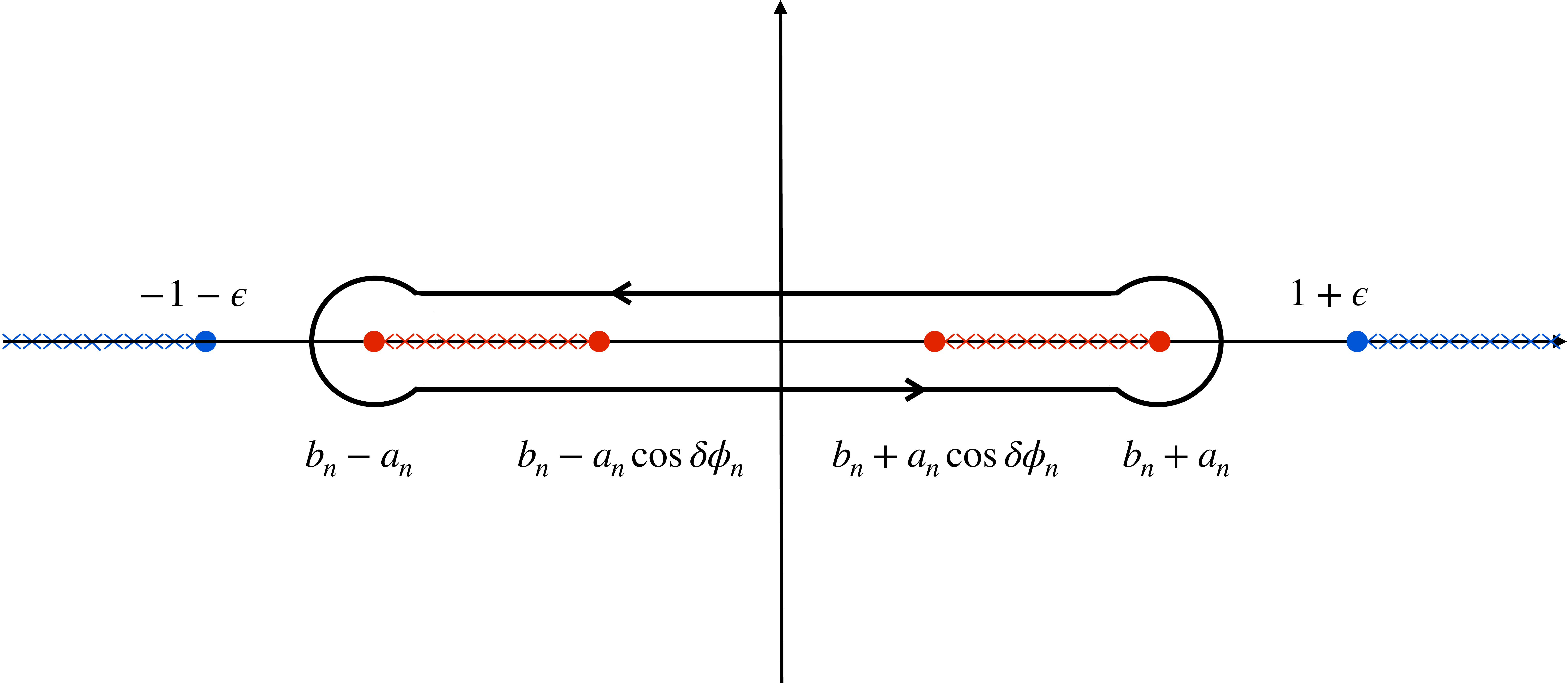}
	\caption{Contour of integration and cuts of the integrand in Eq.~\eqref{eq: contour integral}. The cuts from $\pm(1+\epsilon)$ to the infinity correspond to $ds_\nu(1+\epsilon,z)/dz$ while the cuts inside the contour, $[b_n-a_n,b_n-a_n\cos\delta\phi]$ and $[b_n+a_n\cos\delta\phi,b_n+a_n,]$, are due to the other factor of the integrand.}
	\label{fig: contour2}
\end{figure}

The logarithmic contribution to the entanglement scaling can be computed by considering the discontinuity coefficients. Here, we present their calculation in the general situation in which $G_k$ is discontinuous at a generic mode $k = 2\pi n/N$. We start from the definition~\eqref{eq: b coefficients} of the $b_k$ coefficients corresponding to each discontinuity. First of all, we consider the matrix
\begin{align}
	M_k = (z\mathbb{I}-G_{k}^{-})(z\mathbb{I}-G_k^{+})^{-1},
\end{align}
where $G_k^{\pm} = \lim_{p\to k^\pm}G_p$. The eigenvalues $\mu_k^\pm(z)$ of this matrix can be written in the form
\begin{align}
	\mu_k^\pm(z) = \left(\frac{\sqrt{(b_k-z)^2-a_k^2\cos^2(\delta\phi_{k}/2)}\pm a_k\sin(\delta\phi_{k}/2)}{\sqrt{(b_k-z)^2-a_k^2}}\right)^2,
\end{align}
with $\delta\phi_{k} = \phi_{k}^+-\phi_{k}^-$. Notice also that we have $\mu_k^+(z) = 1/\mu_k^-(z)$, therefore
\begin{align}
	b_k(z) &= \frac{1}{2\pi^2}\left(\ln\mu_k^+(z)\right)^2
	\\\notag
	&=\frac{2}{\pi^2}\left[\ln\left(\frac{\sqrt{(b_k-z)^2-a_k^2\cos^2(\delta\phi_{k}/2)}+a_k\sin(\delta\phi_{k}/2)}{\sqrt{(b_k-z)^2-a_k^2}}\right)\right]^2,
\end{align}
From this expression we compute the coefficient $B_\nu^{(k)}$ of the contribution of this discontinuity to the logarithmic term of the Rényi entropy. For this purpose we plug $b_k(z)$ into the contour integral for $S_{\nu,L}$ then, performing an integration by parts, we obtain
\begin{align}
	B_\nu^{(k)} &= \lim_{\epsilon\to 0^+}\oint_{\mathcal{C}}\frac{dz}{2\pi i}s_\nu(1+\epsilon,z)\frac{db_k(z)}{dz}\label{eq: contour integral}
	\\\notag
	&=-\lim_{\epsilon\to 0^+}\oint_{\mathcal{C}}\frac{dz}{2\pi^3 i}\frac{ds_\nu(1+\epsilon,z)}{dz}\left[\ln\left(\frac{\sqrt{(b_k-z)^2-a_k^2\cos^2(\delta\phi_{k}/2)}+a_k\sin(\delta\phi_{k}/2)}{\sqrt{(b_k-z)^2-a_k^2}}\right)\right]^2.
\end{align}
The integral over the contour $\mathcal{C}$ depicted in Fig. \ref{fig: contour2} can be divided into two integrals along curves enclosing respectively the cuts $[b_k-a_k,b_k-a_k\cos\delta\phi_k]$ and $[b_k+a_k,b_k+a_k\cos\delta\phi_k]$, which in turn can be reduced to two real integrals by performing the integration along the cuts taking into account the change in the phase of the logarithm when we go around the branch points $b_k\pm a_k$ and $b_k\pm a_k\cos\delta\phi_k$. On the other hand, we notice that for integer $\nu>1$, $ds_\nu/dz$ is a meromorphic function with poles located at the points of the imaginary axis \cite{Ares2015PRA,Ares2018PRA}
\begin{align}
	z_l = i\tan\frac{\pi(2l-1)}{2\nu},\quad l = 1,\dots,\nu,\quad l\neq\frac{1+\nu}{2},
\end{align}
and that the another factor of the integrand is analytic in the whole region outside the contour $\mathcal{C}$. We can send this contour to infinity and reduce the calculation of $B_\nu$ to the computation of the corresponding residues. In this way, we obtain the explicit expression
\begin{align}
	B_\nu^{(k)} = \frac{1}{\nu-1}\sum_{l = 1}^{\nu} \left[\ln\left(\frac{\sqrt{(b_k-z_l)^2-a_k^2\cos^2(\delta\phi_k/2)}+a_k\sin(\delta\phi_k/2)}{\sqrt{(b_k-z_l)^2-a_k^2}}\right)\right]^2.
\end{align}
This general formula can be specified in the different cases considered in the main text. In particular in weak long-range case, $1<\alpha_{1},\alpha_{2}<2$, the ground state corresponds to the Bogoliubov vacuum, therefore $f_k = 0$, $a_k = 1$ and $b_k = 0$, $\forall k$. Accordingly, the first term of the expansion vanishes. Moreover the matrix symbol is continuous for generic values of the chemical potential leading to an $\mathcal{O}(1)$ entanglement. The only discontinuities arise at the two quantum critical points $h = h_c = 1,-1+2^{1-\alpha_{1}}$ in correspondence of the critical modes $k = k_c = 0,\pi$. This leads to a logarithmic scaling with coefficient
\begin{align}
	B_\nu^{(k_c)} &= \frac{1}{\nu-1}\sum_{l = 1}^{\nu} \left[\ln\left(\frac{\sqrt{|z_l|^2+\cos^2(\delta\phi_{k_c}/2)}-i\sin(\delta\phi_{k_c}/2)}{\sqrt{|z_l|^2-1}}\right)\right]^2\notag
	\\
	&= \frac{1}{\nu-1}\sum_{l = 1}^{\nu}\left[\arctan\left(\frac{\sin(\delta\phi_{k_c}/2)}{\sqrt{|z_l|^2+\cos^2(\delta\phi_{k_c}/2)}}\right)\right]^2,
\end{align}
where in the last step we have used the identity $\arctan(x) = i[\ln(i+x)-\ln(i-x)]/2$ in order to make the expression of the coefficient explicitly real. The value of $\delta \phi_{k_c}$ depends on the critical point considered and the relative order of the power law decaying exponents $\alpha_{1}$ and $\alpha_{2}$. In particular for $h = 1$ and $k = 0$ we find
\begin{align}
	\delta\phi_{0} = \begin{cases}
		0         &\mathrm{if}\quad \alpha_1<\alpha_2
		\\
		\pi(1-\alpha) &\mathrm{if}\quad \alpha_1=\alpha_2 = \alpha
		\\
		 \pi &\mathrm{if}\quad \alpha_1>\alpha_2.		
	\end{cases}
\end{align}
Leading to the coefficients 
\begin{align}
	B_\nu^{0}(h=1) = \begin{cases}
		0         &\mathrm{if}\quad \alpha_1<\alpha_2
		\\
		\frac{1}{\nu-1}\sum_{l = 1}^{\nu}\left[\arctan\left(\frac{\cos(\alpha\pi/2)}{\sqrt{|z_l|^2+\sin^2(\alpha\pi/2)}}\right)\right]^2   &\mathrm{if}\quad \alpha_1=\alpha_2 = \alpha
		\\
		 \frac{\nu+1}{12\nu}&\mathrm{if}\quad \alpha_1>\alpha_2.
	\end{cases}
\end{align}
On the other hand, for  $h = -1+2^{1-\alpha_{1}}$ and $k = \pi$, $\delta\phi_\pi = \pi$ independently from the values of $\alpha_{1}$ and $\alpha_{2}$. This leads to the scaling coefficient
\begin{align}
	B_\nu^{0}(h=-1+2^{1-\alpha_{1}}) = \frac{\nu+1}{12\nu} \quad\forall\, \alpha_1,\alpha_{2}>1. 
\end{align}

In the strong-long range regime $0<\alpha_{1},\alpha_{2}<1$, the quasiparticle spectrum is discrete also in the thermodynamic limit, this formally leads to an infinite number of discontinuities for any mode $k = 2\pi n/N$, which are labeled by the integer $n = -N/2,\dots N/2$. In particular whenever $\alpha_{1,2}>0$ or $\alpha_{1} = \alpha_{2} = 0$ and $h \neq 0$, the many-body ground state is still the Bogoliubov vacuum characterized by $f_k = 0$, $\forall k$. Accordingly, the matrix symbol in the thermodynamic limit takes the form in Eq.~\eqref{eq: Gn}. The coefficients of the logarithmic scaling is then given by the sum of the contributions coming from all the discontinuity, i.e.,
\begin{align}
	B_\nu = \sum_{n = -N/2}^{N/2}B_\nu^{(n)},
\end{align}
where
\begin{align}
	B_\nu^{(n)} &= \frac{1}{\nu-1}\sum_{l = 1}^{\nu} \left[\ln\left(\frac{\sqrt{|z_l|^2+\cos^2(\delta\phi_{n}/2)}-i\sin(\delta\phi_{n}/2)}{\sqrt{|z_l|^2-1}}\right)\right]^2\notag
	\\
	&= \frac{1}{\nu-1}\sum_{l = 1}^{\nu}\left[\arctan\left(\frac{\sin(\delta\phi_{n}/2)}{\sqrt{|z_l|^2+\cos^2(\delta\phi_{n}/2)}}\right)\right]^2,
\end{align}
with $\delta\phi_n =\phi_{n+1}-\phi_n$. 

Finally, in the mean-field case $\alpha_1 = \alpha_2 = 0$ with zero chemical potential $h = 0$ the quasiparticle spectrum develops an extensive number of degenerate zero modes $\omega_n = 0$ corresponding to all the even modes with $n = 2m$. As a consequence, the ground state is characterized by a finite even mode fermionic population $f_{2m}\neq 0$. The leading order term in the entanglement scaling in this case is then given by the first term of the Fisher-Hartwig expansion corresponding to a volume law. In particular the maximum amount of entanglement allowed by the ground state degeneracy is obtained for $f_{2m} = 1/2$ for every even mode. In this case, the logarithmic corrections become zero since $a_n = b_n = 0$, and therefore  $B_\nu^{(n)}(f_n=1/2) = 0$.  
%%%%

%%%%
\section{Dispersion relation around the critical modes}\label{app: Dispersion relation around the critical modes}
In this Appendix we provide the explicit expression for the Taylor expansion of the quasiparticle spectrum~\eqref{eq: spectrum}, in the weak long-range regime $1<\alpha_{1,2}<2$, at lowest order in $|k - k_c|$, where $k_c = 0$ at
 the critical point $h = 1$, while $k_c = \pi$ at $h = -1+2^{1-\alpha_{1}}$.
In particular, in the proximity of $k = 0$  we find\,\cite{DefenuPRB2019}
\begin{align}
	\tilde{t}_k &= 1+\sin(\alpha_1)\frac{\Gamma(1-\alpha)}{\zeta(\alpha)}|k|^{\alpha_1-1}+O(k^2),
	\\
	\tilde{\Delta}_k &= \sin(\alpha_1)\frac{\Gamma(1-\alpha)}{\zeta(\alpha)}\mathrm{sgn}(k)|k|^{\alpha_2-1}+O(k).
\end{align}
Accordingly, the single particle spectrum takes the form \cite{Solfanelli2022ArXiv}
\begin{align}
	\omega_k = \begin{cases}
		|h-1| + \mathcal{O}(k^{\alpha}-1)                     &\mathrm{if}\quad h\neq 1
		\\
		\mathcal{C}(\alpha)|k|^{\alpha-1}	 + \mathcal{O}(k^{2\alpha-2}) &\mathrm{if}\quad h = 1
	\end{cases},
\end{align}  
where $\alpha = \min\{\alpha_{1},\alpha_{2}\}$, and we have introduced the constant prefactor
\begin{align}
	\mathcal{C}(\alpha) = \begin{cases}
		|\sin(\alpha_1\pi/2)\Gamma(1-\alpha_{1})/\zeta(\alpha_{1})| &\mathrm{if}\quad\alpha_1<\alpha_2
		\\
		|\Gamma(1-\alpha)/\zeta(\alpha)| &\mathrm{if}\quad\alpha_1 = \alpha_2
		\\
		|\cos(\alpha_1\pi/2)\Gamma(1-\alpha_{1})/\zeta(\alpha_{1})| &\mathrm{if}\quad\alpha_1>\alpha_2
	\end{cases}.
\end{align}
On the other hand, near to the $k = \pi$ mode we find \cite{DefenuPRB2019} 
\begin{align}
	\tilde{t}_k &= -1+2^{1-\alpha_1}-\frac{(2^{3-\alpha_1}-1)\zeta(\alpha_1-2)}{2\zeta(\alpha_1)}(\pi-k)^2\notag\\&+ \mathcal{O}((\pi-k)^3),
	\\
	\tilde{\Delta}_k &= \frac{(1-2^{2-\alpha_2})\zeta(\alpha_2-1)}{\zeta(\alpha_2)}(\pi-k)+ \mathcal{O}((\pi-k)^3).
\end{align}
Leading to the $\alpha_{1,2}$ independent dispersion relation
\begin{align}
	\omega_k = \begin{cases}
		|h+1-2^{1-\alpha_1}| + \mathcal{O}((k-\pi)^2)                     &\mathrm{if}\quad h\neq -1+2^{1-\alpha_1}
		\\
		\mathcal{K}(\alpha_2)|\pi-k|+ \mathcal{O}((k-\pi)^3) &\mathrm{if}\quad h = -1+ 2^{1-\alpha_1}
	\end{cases},
\end{align} 
where $\mathcal{K}(\alpha_2) = (1-2^{2-\alpha_{2}})\zeta(\alpha_2-1)/\zeta(\alpha_2)$, $\forall \alpha_1,\alpha_2>1$.

%%%

%%%%
\section{Discontinuities in the strong long-range regime}\label{app: Discontunities in the strong long-range regime}
%
%\begin{figure*}
%	\centering
%	\includegraphics[width=\linewidth]{EvenOdd.pdf}
%	\caption{}
%	\label{fig: EvenOdd}
%\end{figure*}
%
In this Appendix we provide a detailed analysis of the discontinuities of the matrix symbol $G_k$ in the strong long-range regime $0<\alpha_{1},\alpha_{2}<1$ for different values of the chemical potential. As discussed in the Section \ref{sec: Strong long-range regime} of the main text, in this regime the matrix symbol formally develops and infinite number of discontinuities which originate from the discrete nature of the quasiparticle spetrum. However, it is important to notice that, even if the spectrum is labeled by the discrete index $n$ leading to a finite gap between the ground state and the first excited levels, still for $n\gg 1$ all the modes accumulate around $\omega_\infty = |h|$. This means that an extensive number of single-particle states is almost degenerate. Consequently, as long as $h\neq 0$, we may expect only the first few modes around $n = 0$ to provide a significant contribution to the symbol discontinuity, leading to a coefficient $B_\nu(h\neq 0) =\mathcal{O}(1)$. Then, in order to understand the qualitative behavior of  $B_\nu(h\neq 0)$, it is useful to consider the approximation in which only the first discontinuities between the $n = 0$ and the first two degenerate levels $n = \pm 1$ are considered
\begin{align}
	B_\nu(h\neq 0)\approx B_\nu^{(0)}+B_\nu^{(-1)}.
\end{align}
In order to compute this two contributions we have to compute the angles $\phi_{0}$ and $\phi_{\pm 1}$ defined by the conditions
\begin{align}
	\cos\phi_n = \frac{2(h-\tilde{t}_n)}{\omega_{n}},\quad \sin\phi_{n} = -\frac{2\tilde{\Delta}_n}{\omega_{n}}.
\end{align}
For $n = 0$ we find that, independently of the value of $\alpha$, the angle reads
\begin{align}
	\cos\phi_{0} = \begin{cases}
		-1 &\mathrm{if}\quad h<1\\
		0 &\mathrm{if}\quad h>1
	\end{cases},\quad \phi_{0} = \begin{cases}
	\pi &\mathrm{if}\quad h<1\\
	0 &\mathrm{if}\quad h>1.
\end{cases}
\end{align}
This discontinuity at the quantum critical point $h = 1$ is due to the fact that at this point the spectrum becomes gapless for $n = 0$, and it is at the origin of the discontinuity in the scaling coefficient which can be seen in Fig. \ref{fig: StrongLR}a of the main text. The angles for $n = \pm 1$ cannot be computed exactly in close form for generic power law decaying exponent, however as a consequence of the fact that $\tilde{t}_n = \tilde{t}_{-n}$, $\omega_{n} = \omega_{-n}$ while $\tilde{\Delta}_n = -\tilde{\Delta}_{-n}$, we have that 
\begin{align}
	\cos\phi_n = \cos\phi_{-n}\quad \sin\phi_n = -\sin\phi_{-n},
\end{align}
and then $\phi_{n} = -\phi_{-n}$. Combining these properties with Eq.\eqref{eq: B strong long-range} we obtain
\begin{align}
	B_\nu^{(0)} = B_\nu^{(-1)} = 
		\frac{1}{\pi^2(\nu-1)}\sum_{l = 1}^\nu\arctan^2\left[\frac{\cos(\phi_1/2)}{1+\sin^2(\phi_1/2)}\right]  \quad&\mathrm{if}\quad h<1,\\
	B_\nu^{(0)} = B_\nu^{(-1)} = 	\frac{1}{\pi^2(\nu-1)}\sum_{l = 1}^\nu\arctan^2\left[\frac{\sin(\phi_1/2)}{1+\cos^2(\phi_1/2)}\right]  \quad&\mathrm{if}\quad h>1.
\end{align}
\begin{figure*}
	\centering
	\includegraphics[width=\linewidth]{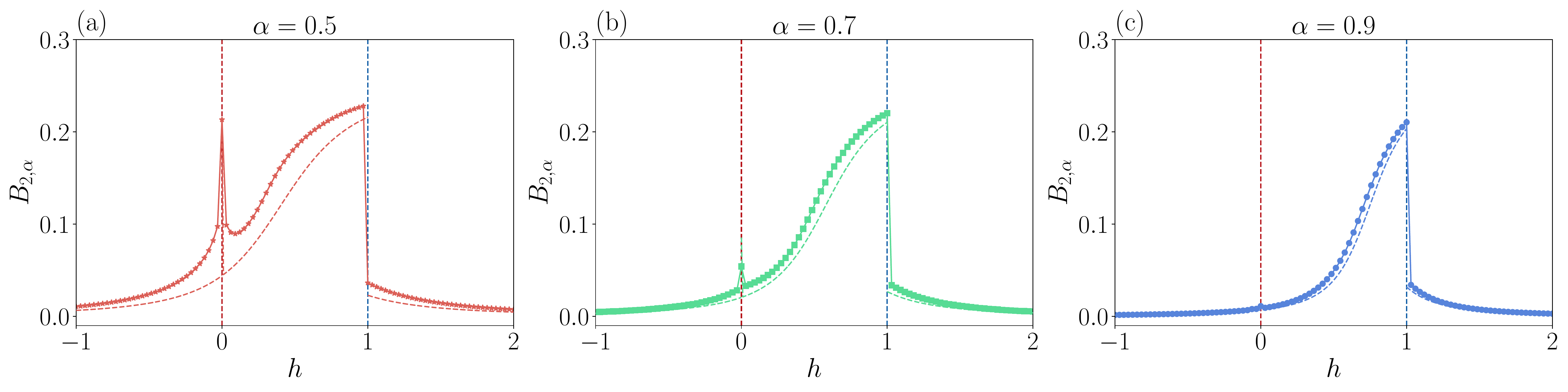}
	\caption{Comparison between the exact values of the logarithmic scaling coefficients of the Rényi-2 entropy, and the single discontinuity approximation (dashed lines) results. The coefficients are plotted as function of the chemical potential $h$ for different values of the decay exponent $\alpha$. }
	\label{fig: SingleDiscApprox}
\end{figure*}
Figure \ref{fig: SingleDiscApprox} shows the comparison between exact values of the logarithmic scaling coefficients of the Rényi-2 entropy $B_{2}$, computed considering the contribution of a formally extensive number of discontinuities (see Eq.~\eqref{eq: B strong long-range}), and the results obtained in the single discontinuity approximation. We notice that the single discontinuity approximation correctly reproduces the qualitative behavior of the scaling coefficients for sufficiently high $\alpha>0.5$ and for values of the chemical potential $h$ which are sufficiently far from $h=0$. In particular, the discontinuity of the coefficients at the quantum critical point $h = 1$ is captured by the approximated result. 

On the other hand, when the chemical potential approaches the $h\to 0$ limit and for sufficiently small decay exponents $\alpha<1/2$, the single discontinuity approximation turns out to be no more accurate. Indeed, in this case the number of relevant discontinuities grows with the subsystem size, leading to a subvolume law entanglement scaling. This fact can be understood by considering the $h = 0$ point. In this case, the spectrum accumulation point becomes $\omega_\infty = 0$. More precisely, it is important to notice that, while at the leading order  as $n\to \infty$ the spectrum goes to zero as $\omega_{n} = \mathcal{O}(n^{\alpha-1})$, independently of the parity of the mode, on the contrary next to leading order corrections differ if $n$ is even or odd.  In particular, if we perform a next to leading order expansion of the terms entering the coefficient $B_2^{(m)}$ (see Eq.~\eqref{eq: B2n}), corresponding to the discontinuity between the  modes $m =2n$ and $m+1 = 2n+1$, we find
%
%\begin{align}
%	t_{2n+1}t_{2n} &= 
%		s_\alpha^{2}n^{2\alpha-2} +\mathcal{O}(n^{2\alpha-3}),\\
%	\Delta_{2n+1}\Delta_{2n} &= 
%		c_\alpha^2 n^{2\alpha-2}-a_\alpha^2 n^{-2}(1-\alpha)/(4\pi^2)+\mathcal{O}(n^{2\alpha-3})\\
%		\omega_{2n+1}\omega_{2n} &=
%\end{align}
%
%
\begin{align}
	\tilde{t}_{2n+1}\tilde{t}_{2n} &= 
		\frac{s_\alpha^{2}}{n^{2-2\alpha}} +\mathcal{O}(n^{2\alpha-3}),\label{eq: todd_teven}\\
	\tilde{\Delta}_{2n+1}\tilde{\Delta}_{2n} &= 
	 	\frac{c_\alpha^2}{n^{2-2\alpha}} -\frac{a_\alpha^2}{n^{2}} +\mathcal{O}(n^{2\alpha-3}),\label{eq: Dodd_Deven}\\
		\omega_{2n+1}\omega_{2n} &= \frac{s_\alpha^{2}+c_\alpha^{2}}{n^{2-2\alpha}}+\frac{b_\alpha}{n^{2}}+\mathcal{O}(n^{2\alpha-3}),\label{eq: wodd_weven}
\end{align}
where we have introduced the expansion coefficients 
\begin{align}
	s_\alpha &= \sin(\alpha\pi/2)\Gamma(2-\alpha)(2\pi)^{\alpha-1},\notag\\
	c_\alpha &= \cos(\alpha\pi/2)\Gamma(2-\alpha)(2\pi)^{\alpha-1},\notag\\
	a_\alpha &= (1-\alpha)/(2\pi),\notag\\
	b_\alpha &=a_\alpha^2(1/2-\cos^2(\alpha\pi/2)) = a_\alpha^2\cos(\alpha\pi)/2.
\end{align}
%%%% 
Now, inserting the large $n$ expansions of Eqs.~\eqref{eq: todd_teven},~\eqref{eq: Dodd_Deven} and~\eqref{eq: wodd_weven} into Eq.~\eqref{eq: B2n}, we see that the denominator is always of order $\mathcal{O}(n^{2\alpha-2})$, while in the numerator the leading order cancels out and we are left with a contribution of order $\mathcal{O}(n^{-2})$ if $\alpha<1/2$ or $\mathcal{O}(n^{2\alpha-3})$ if $\alpha>1/2$. Finally, putting everything together and summing over all the modes we obtain
\begin{align}
	B_{\nu}(h = 0) = \sum_{n}B_{\nu}^{(n)} = \begin{cases} \sum_n{\mathcal{O}(n^{-2\alpha})} = \mathcal{O}(L^{1-2\alpha}) &\alpha<1/2\\
		\sum_n{\mathcal{O}(n^{-1})} = \mathcal{O}(1) &\alpha>1/2
	\end{cases}.
\end{align}
This result leads to the scaling of the Rényi entropy in Eq.~\eqref{eq: entanglement scaling strong LR h0} of the main text.
%\newpage

\providecommand{\href}[2]{#2}\begingroup\raggedright\endgroup
\end{document}